\documentclass{aa}
\usepackage[varg]{txfonts}
\usepackage[caption = false]{subfig}
\usepackage{tabularx}
\usepackage{graphicx}
\usepackage{amsmath}
\usepackage{amssymb}
\usepackage{color}
\usepackage[colorinlistoftodos]{todonotes}
\usepackage{bm}
\usepackage{float}

\usepackage{natbib,twoopt}
\usepackage[breaklinks=true]{hyperref} 
\bibpunct{(}{)}{;}{a}{}{,}             
\makeatletter
  \newcommandtwoopt{\citeads}[3][][]{\href{http://adsabs.harvard.edu/abs/#3}%
    {\def\hyper@linkstart##1##2{}%
     \let\hyper@linkend\@empty\citealp[#1][#2]{#3}}}
  \newcommandtwoopt{\citepads}[3][][]{\href{http://adsabs.harvard.edu/abs/#3}%
    {\def\hyper@linkstart##1##2{}%
     \let\hyper@linkend\@empty\citep[#1][#2]{#3}}}
  \newcommandtwoopt{\citetads}[3][][]{\href{http://adsabs.harvard.edu/abs/#3}%
    {\def\hyper@linkstart##1##2{}%
     \let\hyper@linkend\@empty\citet[#1][#2]{#3}}}
  \newcommandtwoopt{\citeyearads}[3][][]%
    {\href{http://adsabs.harvard.edu/abs/#3}
    {\def\hyper@linkstart##1##2{}%
     \let\hyper@linkend\@empty\citeyear[#1][#2]{#3}}}
\makeatother

\begin{document}

\title{Exposing the plural nature of molecular clouds}
\subtitle{Extracting filaments and the CIB against the true scale-free interstellar medium}

\author{J.-F.~Robitaille\inst{\ref{inst1}}, F.~Motte\inst{\ref{inst1}}, N.~Schneider\inst{\ref{inst2}}, D. Elia\inst{\ref{inst3}}, S. Bontemps\inst{\ref{inst4}}}

\institute{Univ. Grenoble Alpes, CNRS, IPAG, 38000 Grenoble, France,~\email{jean-francois.robitaille@univ-grenoble-alpes.fr}\label{inst1}
		\and
		I. Physik. Institut, University of Cologne, Zülpicher Str.77, 50937 Cologne, Germany\label{inst2}
		\and
		Istituto di Astrofisica e Planetologia Spaziali, INAF, Via Fosso del Cavaliere 100, I-00133 Roma, Italy\label{inst3}
		\and
		OASU/LAB Univ. Bordeaux, CNRS, UMR5804, 33615 Pessac, France\label{inst4}} 

\date{\today\\Submitted to A\&A}

\abstract{We present the Multiscale non-Gaussian Segmentation (MnGSeg) analysis technique. This wavelet based method combines the analysis of the probability distribution function (PDF) of map fluctuations as a function of spatial scales and the power spectrum analysis of a map. This technique allows us to extract the non-Gaussianities identified in the multiscaled PDFs usually associated with turbulence intermittency and to spatially reconstruct the Gaussian and the non-Gaussian component of the map. This new technique can be applied on any data set. In the present paper, it is applied on a \emph{Herschel} column density map of the Polaris flare cloud. The first component has by construction a self-similar fractal geometry as the one produced by fractional Brownian motion (fBm) simulations. The second component is called the coherent component, by opposition to fractal, and includes a network of filamentary structures which demonstrates a spatial hierarchical scaling, i.e. filaments inside filaments. The power spectrum analysis of both components proves that the Fourier power spectrum of the initial map is dominated by the power of the coherent filamentary structures across almost all spatial scales. The coherent structures contribute progressively, more and more from large to smaller scales, without producing any break in the inertial range. We suggest that this behaviour is induced, at least partly, by inertial-range intermittency, a well known phenomenon for turbulent flows. We also demonstrate that the MnGSeg technique is itself a very sensitive signal analysis technique, which allows the extraction of the cosmic infrared background (CIB) signal present in the Polaris flare submillimeter observations and the detection of a characteristic scale for $0.1 \lesssim l \lesssim 0.3$ pc whose origin could partly be the transition of regimes dominated by incompressible turbulence versus compressible modes and other physical processes, such as gravity.}

\maketitle 

\section{Introduction}

A good statistical characterisation and morphological analysis of the interstellar medium (ISM) is important for many astrophysical studies. Identifying the general gas density distribution of molecular clouds as a function of their hosted star formation activity indeed allows us to recognise the dominating physical process of the region and thus to make a link between the ISM structure formation and the emergence of star formation activity. A detailed decomposition of the signal received at different wavelength is also fundamental in order to characterise correctly the properties of the different foreground Galactic components, such as the temperature and column density of ISM gas, and the extragalactic components, such as the cosmic infrared background and the cosmic microwave background.

For these reasons, a reliable morphological analysis of interstellar maps is needed. During the last decades, some statistical tools became the foundations of many theories of ISM structure formation and star formation, as the Fourier power spectrum \citepads{1983A&A...122..282C, 1993MNRAS.262..327G, 2003A&A...411..109M}, the probability distribution function \citep[PDF;][]{1997MNRAS.288..145P, 2008ApJ...688L..79F, 2013ApJ...766L..17S, 2017ApJ...834L...1B} and the $\Delta$-variance \citepads{1998A&A...336..697S, 2008A&A...485..917O, 2008A&A...485..719O}, a method that filters and averages the structures of different sizes $l$ in a map to produce a spectrum showing the relative amount of structure as a function of the structure size. Column density PDFs and the $\Delta$-variance slope have proven to be strongly dependent on the type of forcing, compressive or non-compressive, present in turbulent medium, as well as being sensitive to turbulence intermittency \citepads{2009ApJ...692..364F, 2010A&A...512A..81F}. From theory and molecular cloud simulations, it is proposed that turbulent motions are the main cloud-shaping mechanism and produce a lognormal low (column)-density PDF (e.g. \citeads{2002ApJ...576..870P, 2008ApJ...684..395H, 2014MNRAS.445.1575W, 2018ApJ...863..118B}) followed by a power-law tail due to self-gravitating gas (e.g. \citeads{2011ApJ...727L..20K, 2014ApJ...781...91G}). This scenario is supported by observations using Herschel dust column density maps or extinction maps (e.g. \citeads{2013ApJ...766L..17S, 2015A&A...575A..79S, 2013A&A...553L...8K, 2016MNRAS.461...22P, 2017A&A...606L...2A}) while a pressure governed power-law tail is proposed by \citetads{2011A&A...530A..64K} and \citetads{2014A&A...564A.106T}.

In addition, fluctuations of some physical properties in the ISM, such as the density, can be so large that the average value provides inadequate information. For instance, in their analysis of the statistical properties of the line centroid velocity in a turbulent, compressible but gravitionless simulations, \citetads{1996ApJ...463..623L} found that the global PDF of centroid velocity is close to a Gaussian. However, PDFs of the centroid velocity increments, i.e. the two point statistics for a centroid velocity map separated by a distance $l$, show non-Gaussian wings increasing towards small values of $l$. The same behaviour has also been measured on CO data for the Polaris and the Taurus fields \citepads{2008A&A...481..367H}.

The dilution effect of averaging large density or velocity fluctuations over a field is also affecting the Fourier power spectrum and the $\Delta$-variance analysis. As noted by \citetads{2017MNRAS.466.2529P}, even if the distributions of structure widths in a field, in simulated data or in observations, demonstrate clearly the existence of a dominating `characteristic scale', the spatial power spectrum analysis still show a unique power law attributed to a scale-free medium. The power spectrum analysis of the Polaris field observed by \textit{Herschel} is a good example \citepads{2010A&A...518L.104M, 2010A&A...518L.103M}, where the field is dominated by highly contrasted structures and where the spatial power spectrum has a single power law without any break pointing at a `characteristic scale'. Similar results were obtained on Polaris using the $\Delta$-variance method \citepads{1998A&A...336..697S}, which shows the same information content than the Fourier power spectrum \citepads{2001A&A...366..636B, 2008A&A...485..917O}. Also, recently, a new technique introduced by \citetads{2019A&A...621A...5O}, which compares the power of isotropic and anisotropic structures, shows that Polaris have an almost scale-free filamentary spectrum. \citetads{2011A&A...529A...1S} and \citetads{2014ApJ...788....3E} concluded by applying the $\Delta$-variance on nearby molecular clouds and regions across the Galactic plane that despite of the presence of characteristic scales, the underlying cloud structure is self-similar. This discrepancy between the common scale-free medium measured in the ISM and the presence of highly contrasted filamentary structures remains a fundamental issue in our understanding of the density distribution of the ISM. From these results one can conclude that the typical power spectrum analysis, which is used to measure the hierarchical nature of the ISM, fails to identify the typical sharp transitions in density and filamentary structures in the ISM. These structures are nevertheless physically important, since they are crucial to the mechanisms of star formation, e.g. \citetads{2001ApJ...548..749E}.

Historically, \citetads{1983A&A...122..282C} and \citetads{1993MNRAS.262..327G} measured the first angular power spectrum of HI emission directly from interferometric data. Even if  \citetads{1993MNRAS.262..327G} admitted that various structures in the ISM, such as sheets and filaments, dominate at multiple spatial scales, the fact that there was no preferred angular scale measured in the HI emission power spectrum was interpreted as a sign that turbulence must play a significant role in the hierarchical structure of the ISM. In order to test the fractal nature of the ISM, \citetads{2001ApJ...548..749E} compared HI emission maps of the Large Magellanic Cloud (LMC) with fractal models made from the inverse Fourier transform of random complex-number noise in the $u$-$v$ plane multiplied by a power law. These models are often called fractional Brownian motion (fBm). The resulting fractal models were also exponentiated in order to reproduce the log-normal PDF usually obtained in simulations of magnetohydrodynamical turbulence. However, even if the fractal model respects a similar one point and two point statistics compared to the LMC, i.e. respectively the PDF and the power spectrum analysis, the model lacks all the usual structures associated with the ISM, such as filaments, holes and shells. Fractal models fail to reproduce the common sharp transition in intensity seen in the ISM. Recently, \citetads{2018MNRAS.481..509E} showed that fBm models are not a good approximation of the ISM, but that multifractal analysis offer rather a more complete characterisation of molecular cloud structures.

In the light of these past studies, it is important to recall some fundamental properties of fully developed turbulence. As it is seen in hydrodynamical simulations, a turbulent field can be described as a superposition of some random distribution, as the one produced by fBm simulations, and a set of localised and coherent structures, which also demonstrate a spatial hierarchical scaling \citepads{1992AnRFM..24..395F}. These coherent structures are sometimes identified as a manifestation of intermittency. \citetads{2010A&A...512A..81F} summarised the signature of intermittency in three manifestations: 1) non-Gaussian wings in density and/or velocity PDFs, 2) anomalous scaling of the higher-order ($p\geqslant4$) structure functions of the velocity field and velocity increments, implying that the statistics are increasingly non-Gaussian at small scales, 3) coherent structures of intense vorticity and of strong shocks. In this paper, intermittency is considered in a broad sense, as irregularities and alternation in the spatial statistical distribution of ISM properties, and more specifically for density fluctuations in the case of the present study. This definition corresponds well to ``the dual nature of molecular clouds'' described in the review of \citet{2004Ap&SS.292...89F}, where the diffuse component, traced by $^{12}$CO ($J=$1--0) line emission, is fractal and highly dynamical and the coherent (as opposed to fractal) component, traced by mid-infrared absorption and submillimetric dust thermal emission, is well described by a network of filaments and dense cores.

As demonstrated by \citetads{2001ApJ...548..749E} for HI emission and \citetads{2007A&A...469..595M} for dust far-infrared emission, if exponentiated fBms succeed to reproduce the non-Gaussian wings of log-normal PDFs, these mock fractal simulations fail to reproduce the typical coherent structures in the ISM. Furthermore, \citetads{2014MNRAS.440.2726R} have shown, by applying for the first time the segmentation method described in the present paper, that exponentiated fBms also fail to reproduce the non-Gaussian wings of PDFs measured as a function of spatial scales. As for the PDF analysis of centroid velocity increments, \citetads{2014MNRAS.440.2726R} showed that dust emission at 250 $\mu$m has also more important non-Gaussian wings towards small spatial scales.

In this paper we present a novel decomposition technique based on complex wavelet power spectrum analysis that we have developed to perform an in-depth analysis of the ISM signal \footnote{The codes and tutorials applied on mock simulations are available at \url{https://github.com/jfrob27/pywavan}}. This new technique can be applied on any data set, e.g. column density and velocity centroid maps. Contrary to the Fourier power spectrum, this new technique is sensitive to the dense and coherent filamentary structures. By merging the multiscale PDF analysis with the power spectrum analysis, the technique succeeds to expose, in the case of density fluctuations, the dual, or even plural, nature of molecular clouds and how the diffuse medium is linked to the dense coherent structures. Since this paper focuses on the transition between the two regimes of non-coherent and coherent structures in the ISM, this in-depth analysis will be performed on the Polaris region which represents the early stages of star formation activity in a molecular cloud. In future works, this analysis technique will be applied on centroid velocity maps and numerical simulations, where the intermittent behaviour of both, the density and velocity field, can be compared.

The paper is organised as follows: an overview of the power spectrum analysis is presented in Section \ref{sec:assumptions}and the wavelet power spectrum and the Multiscale non-Gaussian Segmentation technique (MnGSeg) are presented in Section \ref{sec:wavelet_pow_spec}. In Section \ref{sec:application}, we applied MnGSeg to the Herschel column density image of Polaris located at 150 pc, identified the signature of the cosmic infrared background (CIB) and revealed a characteristic scale. Finally, the results are discussed in Section \ref{sec:discussion} and a conclusion is presented in Section \ref{sec:conclusion}.

\section{The assumptions behind the power spectrum analysis}\label{sec:assumptions}

\begin{figure}
\centering
\includegraphics[width=0.48\textwidth]{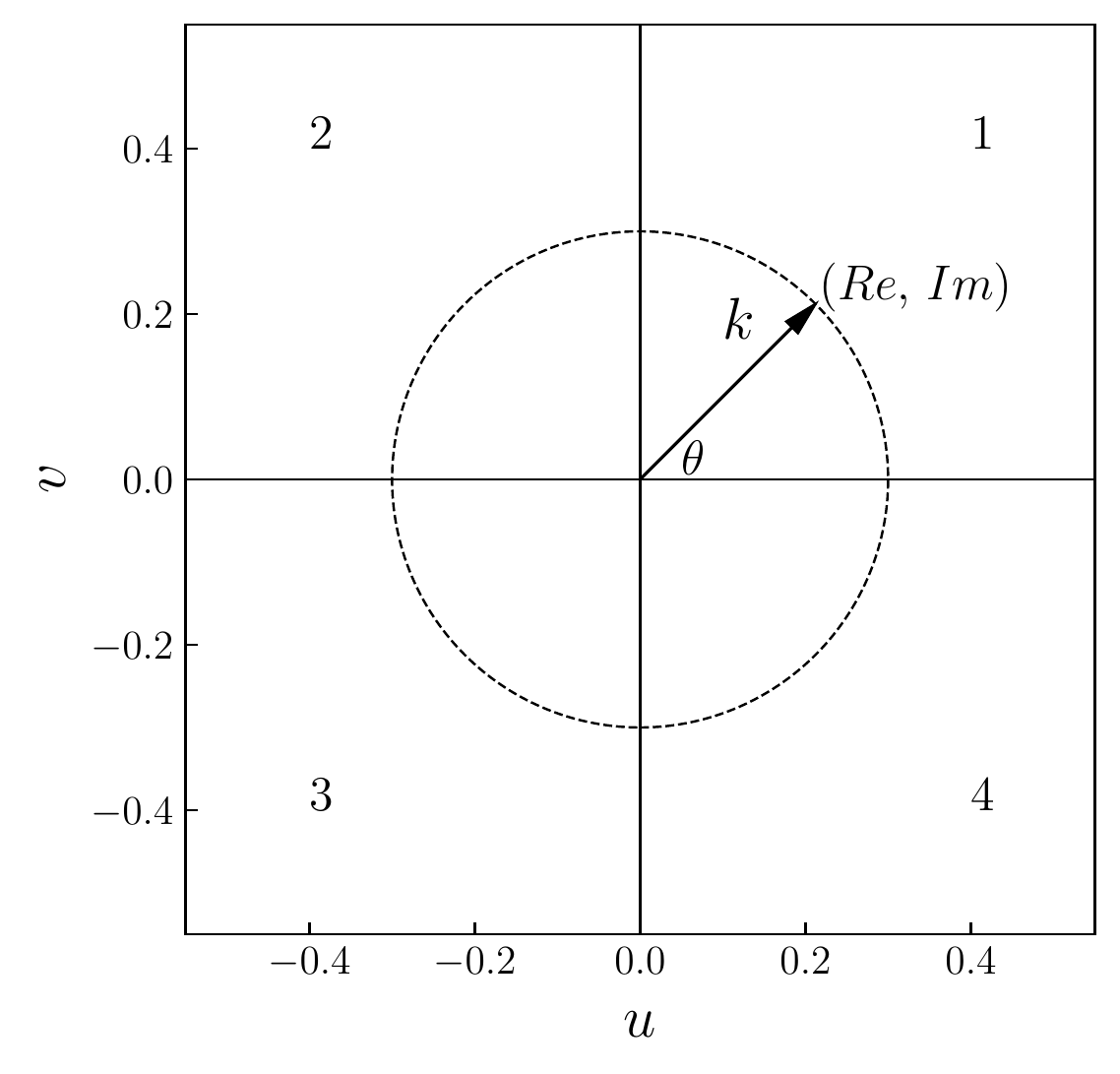}
\caption{Schematic representation of the two-dimensional Fourier space, where $u$ and $v$ are the two dimensions, $\vec{k}$ is the wavenumber and $\theta$ is the azimuthal angle.}
\label{fig:uv_plane}
\end{figure}

The classical power spectrum is usually calculated in the Fourier space. A schematic representation of the two-dimensional Fourier space, or $u$-$v$ plane, is shown in Fig. \ref{fig:uv_plane}. The Fourier transform decomposes the signal $f(\vec{x})$\footnote{Throughout the paper, bold variables denote vectorial quantities. For simplicity, when quantities are averaged over azimuthal angles, as in equation \ref{eq:pow_spec}, the variable is then considered as a scalar.} into a linear combination of Fourier coefficients defined as

\begin{equation}
\hat{f}(\vec{k}) = \int^{\infty}_{-\infty} f(\vec{x}) e^{-2\pi i\vec{x} \cdot \vec{k}}d\vec{k},
\label{eq:fourier_trans}
\end{equation}

\noindent where the wavenumber $\vec{k}$ describes the spatial frequency content of the signal or the image. Each Fourier coefficient is a complex number from which can be calculated the amplitude, $A=\sqrt{Re^2 + Im^2}$, and the phase, $\phi = \arctan(Im/Re)$. The power spectrum analysis of a signal is a statistical measure of the amount of power, $|A|^2$, as a function of the spatial frequency $\vec{k}$. In an ideal world, the experiment leading to the structure formation in the ISM would be reproduced several times under the same initial condition in order to average the different outcomes and to get an adequate statistical sample. This methodology is obviously impossible to achieve in our context. Consequently, we are forced to assume the ergodicity of the medium, so that the local intensity fluctuations averaged over many samples is equal to the spatial average of intensity fluctuations of one realisation. Usually, for the Fourier power spectrum, the information, in the two-dimensional Fourier space, is averaged over the azimuthal angles $\theta$ shown in Fig. \ref{fig:uv_plane}, so that

\begin{equation}
P(k) = \langle |\hat{f}(\vec{k})|^2 \rangle_{\theta}.
\label{eq:pow_spec}
\end{equation}

According to the Kolmogorov theory of turbulence \citepads{1941DoSSR..30..301K}, the velocity power spectrum of an isothermal, subsonic and non-compressive turbulent medium follows a power law over the spatial frequencies:

\begin{equation}
P(k) \propto k^{-\gamma},
\label{eq:pow_spec_propto}
\end{equation}

\noindent where $\gamma$ is the power law index. The power spectrum is equivalent to the Fourier transform of the second order structure function \citepads{2002ApJ...569..841B}. The $p$th-order structure function is defined as

\begin{equation}
\langle |f(\vec{x}')-f(\vec{x})|^p \rangle \propto |\vec{x}' -\vec{x}|^{\zeta_p},
\label{eq:structure_function}
\end{equation}

\noindent where $f(\vec{x})$ referred generally to the fluid velocity $v(\vec{x})$, has a power law of $\zeta_p=p/3$. For the second order longitudinal structure function a power law of $\zeta_2=2/3$, which relates to the Fourier power spectrum index as $\gamma = \zeta_2 +1 = 5/3$. For a three dimensional incompressible and isotropic turbulent medium, the Fourier power spectrum then becomes $P_{\textrm{3D}}(k) \propto k^{-2} k^{-5/3} \propto k^{-11/3}$. Experimental evidence suggests that $\zeta_p$ is smaller than $p/3$ for $p\geqslant4$ \citepads{1994PhRvL..72..336S, 2002ApJ...569..841B, 2004PhRvL..92s1102P}. This implies that the velocity fluctuations are increasingly non-Gaussian at small scales, a phenomenon also referred to as inertial-range intermittency \citepads{1995turb.book.....F}.

The scale-free property, i.e. a scaling function having a unique power law, of the turbulence assumes that the velocity components or the density distribution of a gas dominated by turbulent motions are random variables. Consequently, all the structural properties of a turbulent medium is contained in its power spectrum. The same arguments were used for the development of the $\Delta$-variance analysis \citepads{1998A&A...336..697S}, which shows the same information contained in the power spectrum \citepads{2008A&A...485..917O}. It can be shown that this method is equivalent to the Fourier power spectrum smoothed by the filter spectrum at each scale or size $l$ \citepads{1992AnRFM..24..395F, 2001A&A...366..636B}

Although the Fourier power spectrum is useful to describe the intensity distribution of a map, notably when intensity fluctuations as a function of spatial scales are for a large part random, it fails to provide an accurate description of the distribution of more complex medium. In the case of the ISM gas distribution, where compressive mechanisms and the intermittency of turbulence produced dense filamentary structures, intensity distribution as a function of spatial scales are no longer random, nor isotropic, as it is the case for instance in fractal simulations. These inhomogeneities in the medium also break the ergodic assumption. The spatial average of intensity fluctuations is no longer representative of intensity fluctuations occurring locally in the map.

The Fourier power spectrum loses all the local information associated with intensity fluctuations as a function of the spatial scales. Because the trigonometric functions associated to the Fourier coefficients, $\hat{f}(\vec{k})$ in equation \ref{eq:fourier_trans}, oscillate forever all the information content of $f(\vec{x})$ is completely delocalised \citepads{1992AnRFM..24..395F}. As a solution to this limitation, \citetads{2014MNRAS.440.2726R} showed that the analysis of the wavelet power spectrum of an image allows one to have access not only to the spatial frequency content of the signal, but also to the information on the localised intensity fluctuation in an image as a function of the spatial scales. The gain of information is substantial and can be used to localise the transitions to high-intensity regions, perhaps associated with important changes in the main physical mechanisms at play.

By opposition to the $\Delta$-variance spectrum, which averages all the wavelet coefficients as a function of spatial scale or size $l$, and therefore loses all the local intensity fluctuation information, the Multiscale non-Gaussian segmentation (MnGSeg) technique, as primarily described by \citetads{2014MNRAS.440.2726R}, isolates the random and isotropic component of a map as a function of the spatial scales by analysing the PDF of complex-valued and directional wavelet coefficients before analysing the power spectrum. This method has the advantage to separate the map component which satisfies the ergodic assumption from the dense and anisotropic structures, such as the ubiquitous interstellar coherent filaments, which normally bias the Fourier power spectrum analysis. These two components refer to ``the  dual nature of molecular clouds'' described by \citetads{2004Ap&SS.292...89F}. In addition to the non-biased power spectrum, which can measure the true scale-free nature of a map, such multiscale segmentation technique allows us to separate density structures contributing to the non-Gaussian part of the PDFs, i.e. structures that, in the simple isotherm model, correspond to intermittency, but in the context of the more complex ISM, may also correspond for instance to self-gravitating structures.

The next section reintroduces the procedure described by \citetads{2014MNRAS.440.2726R} and improve it further.

\section{The MnGSeg technique}\label{sec:wavelet_pow_spec}

\subsection{The wavelet power spectrum}

Wavelet transforms are designed to analyse local fluctuations in a signal. The wavelet transform is obtained with the convolution of a map $f(\vec{x})$ with a family of translated and dilated wavelets generated from the mother wavelet function $\psi(\vec{x})$:

\begin{equation}
\tilde{f}(l,\vec{x})= \frac{1}{l} \int \int_{-\infty}^{+\infty} f(\vec{x}')\psi^*\Big(\frac{\vec{x}'-\vec{x}}{l}\Big)d\vec{x}'.
\label{eq:wavelet_transform}
\end{equation}

\noindent As a result, a low or a high wavelet coefficient $\tilde{f}(l,\vec{x})$ means that at the position $\vec{x}$ and spatial scale $l$, the signal has a low or a high variation compared to the mean value of the signal. As for the Fourier transform, the wavelet transform respects the Plancherel relation:

\begin{equation}
\int \int_{-\infty}^{\infty} |f(\vec{x})|^2 d\vec{x} = \frac{2\pi}{C_{\psi}} \int_{0}^{\infty} P(l) \frac{dl}{l},
\label{eq:Plancherel}
\end{equation}

\noindent where

\begin{equation}
P(l)=\int \int_{-\infty}^{\infty} |\tilde{f}(l,\vec{x})|^2 d\vec{x}
\label{eq:wavelet_spectrum}
\end{equation}

\noindent and

\begin{equation}
C_{\psi}=\int \int_{-\infty}^{+\infty} \frac{|\hat{\psi}(\vec{k})|^2}{|\vec{k}|^2}d\vec{k}.
\label{eq:second_condition}
\end{equation}

\begin{figure*}
\centering
\includegraphics[width=1.0\textwidth]{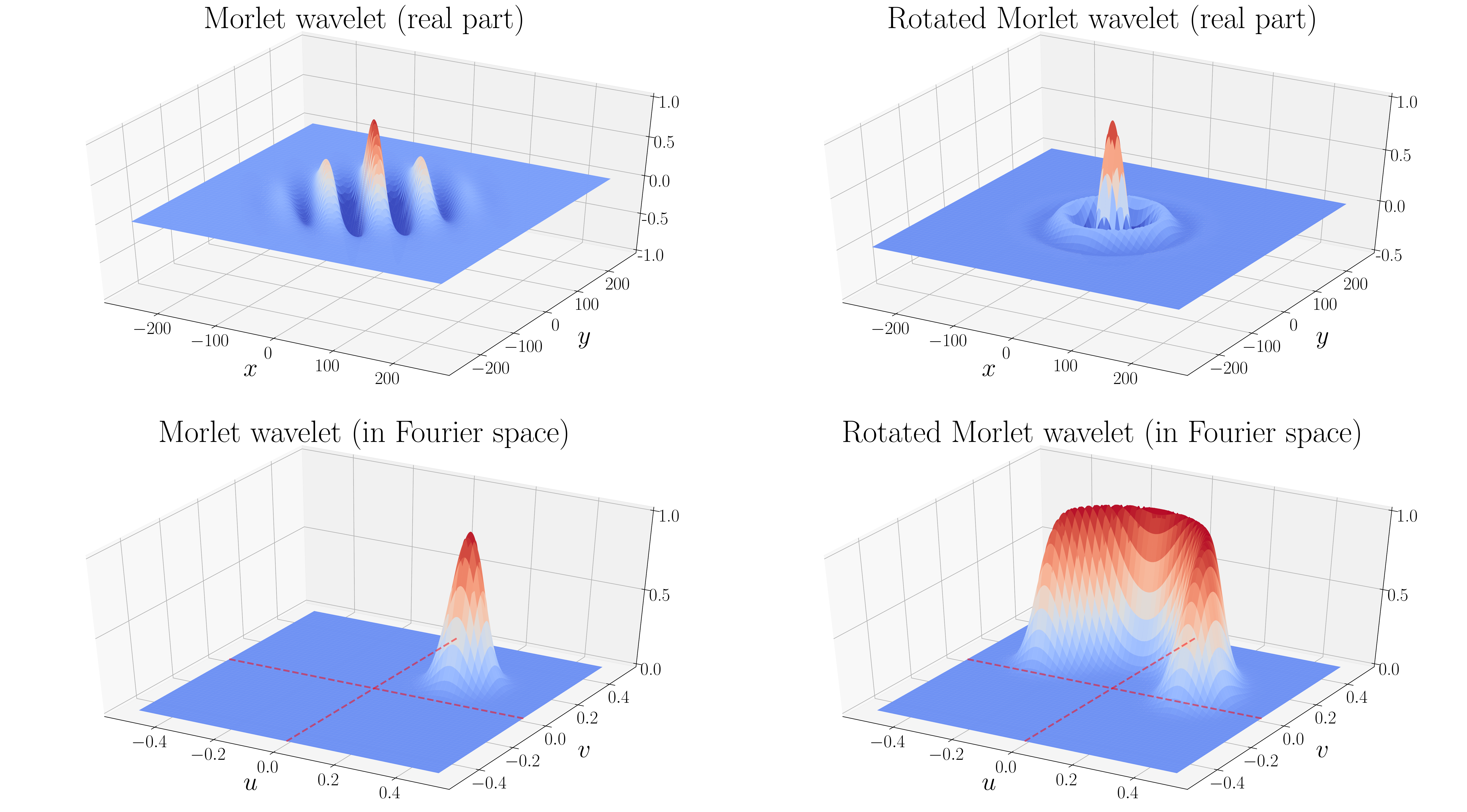}
\caption{Top left panel: The real part of the Morlet wavelet. Top right panel: The Morlet wavelet rotated over an azimuthal angle of $\pi$; also called the Fan wavelet \citepads{2005CG.....31..846K}. Bottom left panel: The Morlet wavelet in the Fourier space. Bottom right panel: The Morlet wavelet rotated over an azimuthal angle of $\pi$ in the Fourier space.}
\label{fig:Morlet}
\end{figure*}

\noindent Equation \ref{eq:wavelet_spectrum} represents the global wavelet power spectrum. This relation is true for all wavelet functions $\psi(x)$. However, because some functions have a better resolution in the frequency space, some wavelets are more accurate to estimate the power spectrum of a signal. \citetads{2005CG.....31..846K} showed that the Morlet wavelet is the best wavelet function to reproduce the Fourier power spectrum. The Morlet wavelet is complex-valued and anisotropic. It is defined in the Fourier space as

\begin{equation}
\begin{split}
\hat{\psi}(\vec{k}) & =\textrm{e}^{-|\vec{k}-\vec{k}_0|^2/2} \\
                                     & =\textrm{e}^{-[(u-|\vec{k}_0|\cos \theta)^2+(v-|\vec{k}_0|\sin \theta)^2]/2},
\end{split}
\label{eq:Morlet_Fourier}
\end{equation}

\noindent where the constant $|\vec{k}_0|$ is set to $\pi\sqrt{2/\ln2} \approx 5.336$ to ensure that the admissibility condition \footnote{The admissibility condition requires the zero mean value of the wavelet function, $\int^{+\infty}_{-\infty}\psi(x)dx=0$. Since, without any correction, $\hat{\psi}(0) \neq 0$ for the Morlet wavelet, has defined in equation \ref{eq:Morlet_Fourier}, this wavelet is only marginally admissible.} is almost met \citepads{2005CG.....31..846K}. As defined in Eq. \ref{eq:Morlet_Fourier}, in the Fourier space the complex Morlet wavelet is equivalent to a Gaussian kernel that can easily sample spatial frequencies as a function of the azimuthal angle $\theta$ (see bottom left panel of Fig. \ref{fig:Morlet}). By opposition to a real isotropic wavelet, complex-valued wavelets with an azimuthal dependency allow one to estimate the true power as defined in Eq. \ref{eq:pow_spec}.

With this additional azimuthal dependency, it is also possible to estimate the power spectrum of an image by integrating over $\theta$ instead of over $\vec{x}$ as in equation \ref{eq:wavelet_spectrum}. Because of the finite azimuthal resolution of the Morlet wavelet, the integration is trade for a discrete summation over a limited number of angles. \citetads{2005CG.....31..846K} showed that the optimal angle interval for an efficient and uniform sampling is $\delta \theta=2\sqrt{-2\ln 0.75}/|\vec{k}_0|$ (see top right panel of Fig. \ref{fig:Morlet}). Equation \ref{eq:wavelet_spectrum} thus becomes:

\begin{equation}
P(l,\vec{x})=\frac{\delta \theta}{N_{\theta}} \sum_{j=0}^{N_{\theta}-1}  |\tilde{f}(l,\vec{x},\theta_j)|^2,
\label{eq:Fan_wavelet}
\end{equation}

\noindent where $\tilde{f}(l,\vec{x},\theta)$ are the Morlet wavelet coefficients for map $f(\vec{x})$ and $N_{\theta}=\Delta \theta/\delta \theta$ is the number of directions $\theta$ needed to sample the Fourier space over the range $\Delta \theta$. Since, for a real image, the quadrants 3 and 4 represented in Fig. \ref{fig:uv_plane} are redundant, they are the complex conjugate of respectively quadrants 1 and 2, only angles in quadrant 1 and 2 need to be sampled, which leads to $\Delta \theta=\pi$ (see bottom right panel of Fig. \ref{fig:Morlet}). The convolution operation for the wavelet transform can be done directly in the Fourier space, so that $\tilde{f}(l,\vec{x},\theta) = \mathcal{F}^{-1}\left\{\hat{f}(\vec{k})\hat{\psi}_{l,\theta}^*(\vec{k})\right\}$, where $\mathcal{F}^{-1}$ denotes the inverse Fourier transform.

Compared to the Fourier power spectrum analysis, the complex Morlet wavelet power spectrum analysis, as defined in Eq. \ref{eq:Fan_wavelet}, is not only dependent of the spatial scale $l$, but also of the map position $\vec{x}$. This property provides a far more complete description of intensity fluctuations as a function of spatial scale in a map. From $P(l,\vec{x})$, one can recover the global wavelet power spectrum by averaging the power over all positions $\vec{x}$,

\begin{equation}
P(l)=\frac{1}{N_{\vec{x}}} \sum_{\vec{x}} P(l,\vec{x}),
\label{eq:global_scalogram}
\end{equation}

\noindent where $N_{\vec{x}}=N_x \times N_y$ corresponds to the number of pixels in the map. By converting the spatial scale $l$ to the Fourier wavenumber $k$ using $k=|\vec{k}_0|/l$, one can compare the global wavelet power spectrum of equation \ref{eq:global_scalogram} directly with the Fourier power spectrum defined in equation \ref{eq:pow_spec}.

\subsection{Non-Gaussian segmentation}\label{sec:segmentation}

The complex wavelet power spectrum allows one to analyse the local variation in intensity and global power as a function of spatial scales. This complete description of the intensity fluctuations in the map can be used to isolate the random component linked to the scale-free nature of the ISM. The residues of this segmentation procedure does not satisfy the randomness and the ergodicity assumptions and are called the \textit{coherent structures} because of their fundamental properties of being spatially correlated across the scales.

As for the previous segmentation analysis by \citetads{2014MNRAS.440.2726R}, in order to separate these two components we use the coherent vorticity extraction algorithm \citep{2012PhyD..241..186N, Azzalini:2005ed}. This iterative algorithm was initially developed as a method to determine the optimal denoising threshold among wavelet coefficients. In many cases, denoising consists in deleting the wavelet coefficients of a noisy signal whose modulus is below a threshold, usually found at small scales, and reconstructing the denoised signal from the remaining coefficients \citep{Azzalini:2005ed}. In our case, the algorithm is applied at every spatial scales and as a function of the azimuthal direction. For our analysis, the noisy random coefficients are considered to be the scale-free component of the map and the other set of coefficients is considered to be the coherent structures of the map. The algorithm is defined as follows, let $\Phi$ be the threshold splitting the non-Gaussian terms from the Gaussian terms in the wavelet coefficient distribution and $\mathbb{L}_{\Phi}$ be the function indicator. The threshold $\Phi$ is first estimated according to the variance,

\begin{equation}
\sigma_{l,\theta}^2(\Phi)=\frac{1}{N_{l,\theta}(\Phi)}\sum_{\vec{x}} \mathbb{L}_{\Phi}( |\tilde{f}_{l,\theta}(\vec{x})|) |\tilde{f}_{l,\theta}(\vec{x})|^2,
\end{equation}

\noindent where

\begin{equation}
\mathbb{L}_{\Phi}(|\tilde{f}_{l,\theta}(\vec{x})|)=\left\{
        									          \begin{array}{ll}
  									          1 & \qquad \mathrm{if} \quad |\tilde{f}_{l,\theta}(\vec{x})| < \Phi \\
 									          0 & \qquad \mathrm{else}. \\
 									         \end{array}
  									         \right.
\end{equation}

\noindent and

\begin{equation}
N_{l,\theta}(\Phi)=\sum_{\vec{x}} \mathbb{L}_{\Phi}( |\tilde{f}_{l,\theta}(\vec{x})|).
\end{equation}

\noindent The iterative calculation then converges to an optimal value of the threshold $\Phi$ which allows one to separate outliers from randomly distributed wavelet coefficients. The sequence is defined by:

\begin{equation}
\left\{
        	 \begin{array}{l}
  	   \Phi_{0}(l,\theta)=\infty \\
 	   \Phi_{n+1}(l,\theta)=q \sigma_{l,\theta}(\Phi_{n}(l,\theta)), \\
 	 \end{array}
  	 \right.
\label{eq:q_parameter}
\end{equation}

\noindent where $q$ is a dimensionless constant controlling how restrictive is the definition of non-Gaussianities. The first study using the MnGSeg technique chose the value of $q$ following two criteria: the normal distribution of the power for the Gaussian features and the unique power-law of its power spectrum. In the present paper, $q$ is dynamical and thus dependent of the spatial scale. We considered that the amount of non-Gaussianities produced by turbulence intermittency and/or compressive physical processes, as shocks, as a function of scale is unknown and that the non-Gaussianities contribution vary from one scale to another. Consequently, in order to adjust the parameter $q$ to its optimal value, at each spatial scale, when the algorithm converges to an optimal value for a threshold $\Phi$, the skewness, the third moment of the distribution, of the Gaussian wavelet coefficient distribution is calculated. If the skewness is larger than $0.4$, the value of $q$ is diminished by $0.1$. This operation is repeated until convergence of the parameter $q$. The value $0.4$ is justified by the fact that the distribution skewness is evaluated on the absolute value of complexe valued wavelet coefficients and that, consequently, the distribution is not Gaussian, but Rician. Rician distributions are not completely symmetrical and have a non-zero skewness.

\subsection{Power spectra analysis and reconstruction}

After the convergence of the extraction algorithm, two sets of wavelet coefficients are obtained, the Gaussian set $\tilde{f}^{\textrm{~G}}_{l,\theta}(\vec{x})$ and the non-Gaussian set, also called the coherent set, $\tilde{f}^{\textrm{~C}}_{l,\theta}(\vec{x})$. Then equations \ref{eq:Fan_wavelet} and \ref{eq:global_scalogram} can be applied in order to calculate the power spectrum of both sets, $P^G(k)$ and $P^C(k)$. Since both power spectra are obtained from the squared amplitude of independent wavelet coefficients, the total power spectrum, equivalent to the Fourier power spectrum is simply the linear combination of the segmented components:

\begin{equation}
P(k) = P^G(k) + P^C(k)
\label{eq:lin_comb}
\end{equation}

\noindent As demonstrated by \citetads{2014MNRAS.440.2726R}, images corresponding to both set of coefficients can also be reconstructed. Originally, the reconstruction formula used the same synthesising wavelet as the analysing wavelet, i.e. the Morlet wavelet. However, thanks to the redundancy of continuous wavelets, many reconstruction formulas exit for a wavelet decomposed signal \citepads{1992AnRFM..24..395F}. J. Morlet found empirically that even the delta function can be used to reconstruct the signal. In that case the reconstruction formula becomes:

\begin{equation}
f(\vec{x}) = C_{\delta} \sum_l \sum_{j=0}^{N_{\theta}-1} l \tilde{f}(l,\vec{x},\theta_j) + \mu_0,
\label{eq:synthesis}
\end{equation}

\noindent where $\mu_0$ is the mean value of the original map and $C_{\delta}$ is a correction factor. The reconstruction was found to be optimal when the scale separation in logarithm, $\Delta \ln l$, is set to $\delta \theta$, the same separation as for the azimuthal angle. This separation allows one to construct a quasi-orthogonal set of wavelet coefficients representing the signal. The reconstruction is not perfect but it is largely sufficient for most applications. The correction factor $C_{\delta}$ is set to $\sigma_r/\sigma_0$, where $\sigma_r$ and $\sigma_0$ are the standard deviations of respectively the reconstructed map and the original map.

\noindent Because of the linearity of Eq. \ref{eq:synthesis} and the linearity of wavelet transforms:

\begin{equation}
f(\vec{x}) = C_{\delta}\left(f^G(\vec{x}) + f^C(\vec{x})\right)+ \mu_0,
\label{eq:lin_comb_spatial}
\end{equation}

\noindent where,

\begin{equation}
f^G(\vec{x}) = \sum_l \sum_{j=0}^{N_{\theta}-1} l \tilde{f}^G(l,\vec{x},\theta_j)
\label{eq:synthesisG}
\end{equation}

\noindent and

\begin{equation}
f^C(\vec{x}) = \sum_l \sum_{j=0}^{N_{\theta}-1} l \tilde{f}^C(l,\vec{x},\theta_j).
\label{eq:synthesisC}
\end{equation}

\noindent $f^G(\vec{x})$ and $f^C(\vec{x})$ are the reconstructed maps of the Gaussian and coherent wavelet coefficient sets, respectively.

\section{Application on Polaris flare column density map}\label{sec:application}

\subsection{Data}

\begin{figure*}
\centering
\includegraphics[width=1.0\textwidth]{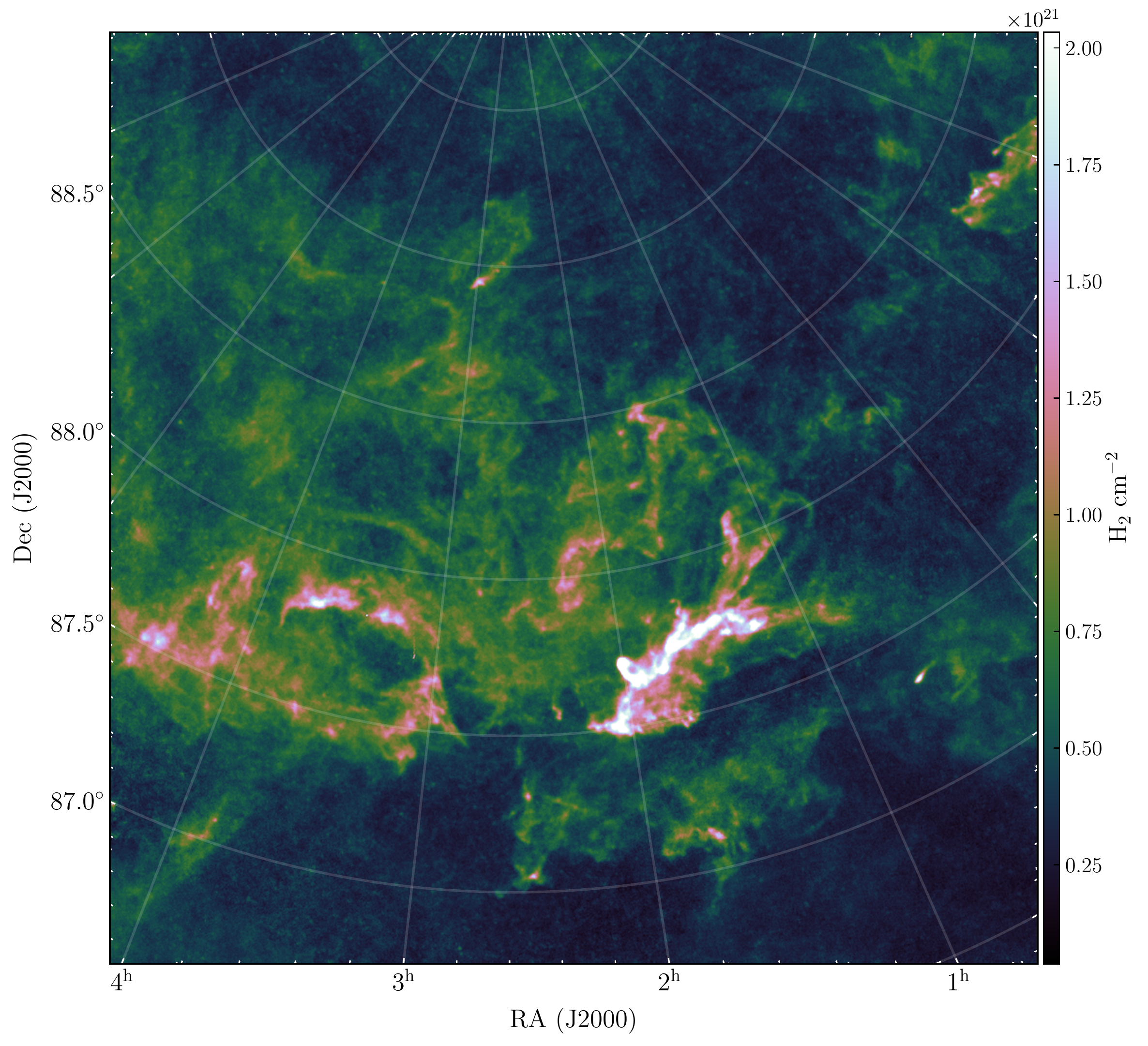}
\caption{Polaris column density map derived from \emph{Herschel} imaging data taken as part of the HGBS and ``Evolution of interstellar dust'' key programs. The brightest structure with the highest column density value located on the south-western part of the field has been labeled as MCLD 123.5+24.9 and as the ``saxophone'' by \citetads{2013ApJ...766L..17S}.}
\label{fig:polaris}
\end{figure*}

We used the SPIRE $250~\mu$m and the column density map of Polaris derived from \emph{Herschel} imaging data taken as part of the HGBS and ``Evolution of interstellar dust'' key programs \citepads{2010A&A...518L.102A, 2010A&A...518L..96A, 2010A&A...518L.104M}. The column density map was produced following a procedure described in most \emph{Herschel} papers (see e.g., \citeads{2013A&A...550A..38P}; \citeads{2015A&A...575A..79S}; \citeads{2015A&A...584A..91K}) and adopting a mean molecular weight per hydrogen molecule $\mu_{\rm H2} = 2.8$ \citepads{2008A&A...487..993K}. The column density map has an angular resolution of $36\arcsec$, corresponding to the half-power beam width resolution of \emph{Herschel}/SPIRE $500~\mu$m data and is estimated to be accurate to better than $\sim50\%$ (e.g., \citeads{2014A&A...562A.138R}). The pixel size is of $14\arcsec$. The $250~\mu$m map has an angular resolution of $18\arcsec$ and a pixel size of $6\arcsec$. The Polaris column density map is shown in Fig. \ref{fig:polaris}.

\subsection{Fourier and wavelet power spectra}

The Fourier and wavelet power spectra have been calculated on the Polaris flare region located at high Galactic latitude ($b \sim 25^{\circ}$). This region has no ongoing star-formation activity. Only prestellar cores and 
unbound starless cores have been detected so far \citepads{2010A&A...518L.102A, 2010A&A...518L..92W}.

The power spectrum analysis of this region has been done by \citetads{2010A&A...518L.104M} on the three wavelengths observed by the SPIRE instrument, 250, 350 and 500 $\mu$m, and the IRAS 100 $\mu$m map. All power spectra, once corrected for the noise, the SPIRE beam and the point sources contribution, show a straight power law with, within the uncertainties, a similar power law of $-2.7$. This measurement also agrees with the previous power law measured by \citetads{1998A&A...336..697S} using the $\Delta$-variance method on this low-density region properly traced by CO.

\begin{figure}
\centering
\includegraphics[width=0.48\textwidth]{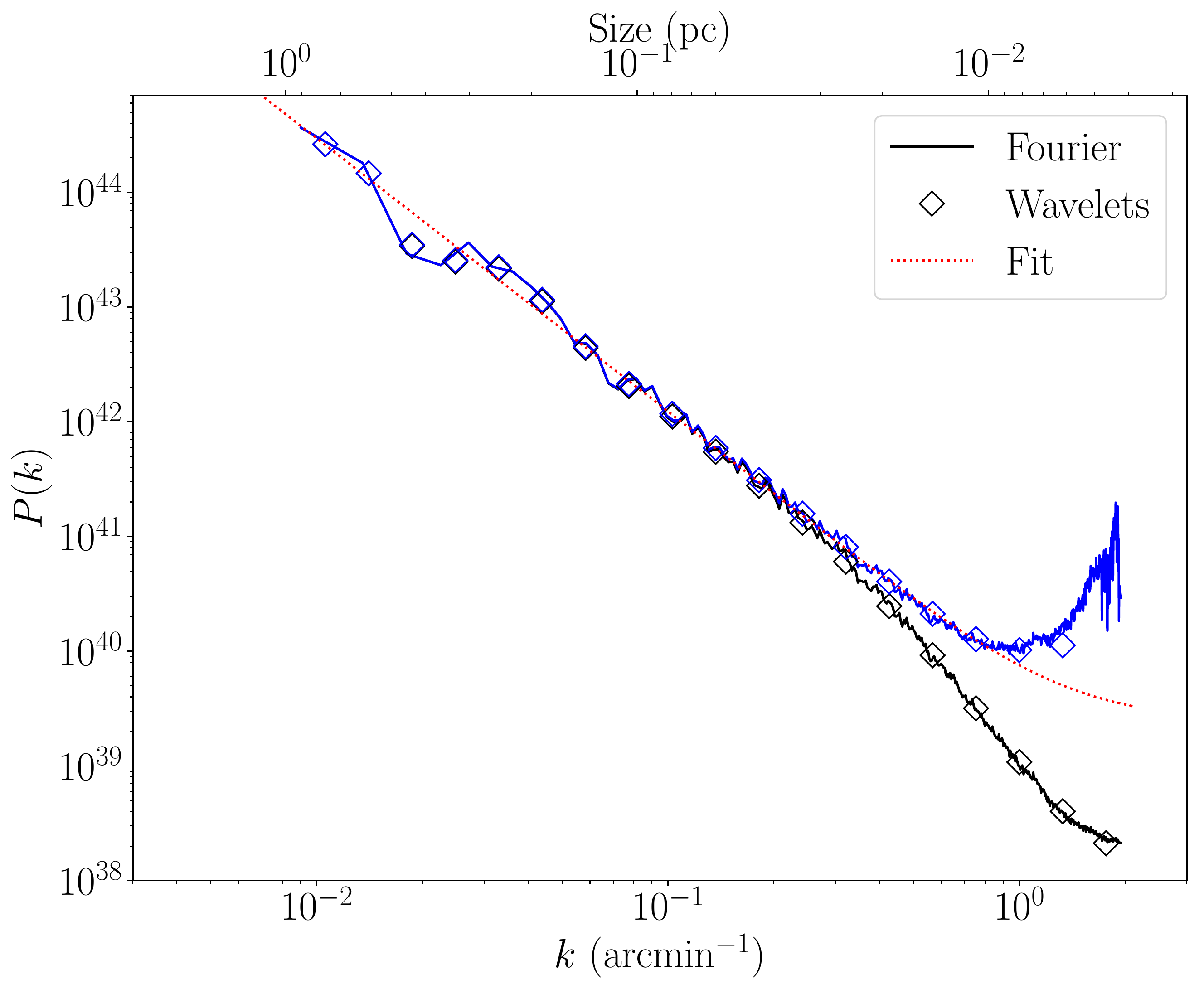}
\caption{The Fourier (solid lines) and wavelet (diamonds) power spectra of the Polaris flare region presented in Fig. \ref{fig:polaris}. The black line and black symboles represent the original power spectrum, $P(k)$ in Eq. \ref{eq:MAMD_model}, while the blue line and blue symboles represent $P_{\textrm{sky}}(k)$ in Eq. \ref{eq:MAMD_Psky}.}
\label{fig:polaris_pow_spec}
\end{figure}

Figure \ref{fig:polaris_pow_spec} presents our power spectrum analysis of the Polaris flare region presented in Fig. \ref{fig:polaris}. To reduce the map edge effects, the Fourier transform and wavelet transforms have been done on a map $\sim 1.25$ times larger than the original one, where the frame pixel values are zeros and an apodisation has been applied over 3\% of the original map edges. The mean pixel value of the map was subtracted prior to the apodisation to reduce the gap between the intensity of the signal and zero value pixel frame. In order to produce the wavelet power spectrum, equations \ref{eq:Fan_wavelet} and \ref{eq:global_scalogram} were calculated on the $\sim 1.25$ times larger map for every scales corresponding to the diamond symbole in Fig. \ref{fig:polaris_pow_spec}. \citetads{2010A&A...518L.104M} previously modelled the Polaris power spectrum as

\begin{equation}
P(k) = \Gamma(k)P_{\textrm{sky}}(k)+N(k)
\label{eq:MAMD_model}
\end{equation}

\noindent where,

\begin{equation}
P_{\textrm{sky}}(k) = A_{\textrm{ISM}}k^{\gamma} + P_{0}.
\label{eq:MAMD_Psky}
\end{equation}

\noindent The factor $\Gamma(k)$ is the telescope transfer function, $N(k)$ is the noise level and $P_{0}$ modelled the excess of power at small scales induced by point sources and the cosmic infrared background (CIB) associated with unresolved infrared galaxies at high redshift. In order to measure the power associated with $P_{\textrm{sky}}(k)$, the original power spectrum is first subtracted by the noise level and then divided by the telescope transfer function. For this analysis, the noise level is estimated by the last point of the wavelet power spectrum at $k \approx 1.75$ arcmin$^{-1}$. The fitted values for the Fourier power spectrum are listed in Table \ref{tab:Fourier_pow_spect_fit}. The fit has been estimated between 0.05 and 0.8 arcmin$^{-1}$. The measured power law is shallower than the previous measurements made on individual wavelength maps by \citetads{2010A&A...518L.104M} (a corresponding power spectrum analysis of the \emph{Herschel} 250 $\mu$m map is done in section \ref{sec:CIB}).

\begin{table}
\centering
\small
\caption{Fit values for the column density map Fourier power spectrum}
\label{tab:Fourier_pow_spect_fit} 
\begin{tabular}{lccc}
\hline \hline
&$A_{\textrm{ISM}}$ & Power-Law $(\gamma)$ & $P_0$ \\
&(H$_2$ cm$^{-2}$)$^2$ & & (H$_2$ cm$^{-2}$)$^2$ \\ \hline

 & $(5.1\pm0.2) \times 10^{39}$ & $2.38\pm0.02$ & $(2.5\pm0.5) \times 10^{39}$  \\

\hline
\end{tabular}
\end{table}

The wavelet power spectrum matchs very well with the Fourier power spectrum except for small deviations noticeable near the noise level.

\subsection{Intermittency}\label{sec:intermittency}

The good correspondence between the Fourier and wavelet power spectra validates equation \ref{eq:global_scalogram} and suggests that the Fourier power spectrum is sensitive only to the mean variation of column density over the map as a function of spatial scales. However, as it is seen in many numerical simulations and as it is generally measured in column density PDFs of star formation regions, molecular cloud dense structures produce a tail at large densities on the column density distribution. These tails do not generally have a significant impact on the mean value of a statistical distribution. Large skewness has also been predicted as a function of spatial scales on centroid velocity increments by \citet[][their Fig. 9]{2010A&A...512A..81F} for solenoidal and compressive forcing. PDFs should be close to Gaussian distributions (in semi-log plots) at large scales and present exponential tails at smaller scales. Concerning Polaris, they concluded that the kurtosis measured on centroid velocity increments (CVIs) were compatible with intermittency of solenoidal (incompressible) forcing. This measure of intermittency was in good agreement with the CVIs of $^{12}$CO(1--0) IRAM map by \citet[][their Fig. 4]{2008A&A...481..367H}.

In this paper, we use the spatial scale filtering property of wavelet transforms as an alternative to the two-point statistics used to calculate CVIs. Besides, the spatial scale filtering is performed on a column density map rather than on a velocity centroid map. This choice allows us to perform the multiscale analysis on a wider range of spatial scales and to investigate the intermittent behaviour of the density field.

We recall that in the case of compressible and supersonic turbulence, the velocity field and density fluctuations become strongly coupled \citepads{2007ApJ...665..416K}. Large velocity-shears indeed produce intermittent structures in the velocity field that can follow boundaries of high density structures traced in a column density map \citepads{2009A&A...500L..29H}. Also, \citetads{2010A&A...512A..81F} showed that the PDFs of the logarithm of the density are usually roughly consistent with log-normal distributions for both solenoidal and compressive forcings. They attributed the non-Gaussian higher-order moments deviations of the distributions, such as the skewness and the kurtosis, to turbulence intermittency. They suggested that these deviations can be caused by head-on collisions of strong shocks or oscillations in very low-density rarefaction waves. In agreement with this interpretation, \citetads{2008A&A...481..367H} showed that for Polaris the centroid velocity structures associated with increments of CVI tails closely follow the boundaries of the optically-thin $^{12}$CO emission traced by the broad $^{12}$CO line wings. Finally, \citetads{2003A&A...411..109M} showed that power spectra of integrated emission and centroid velocity fields of the high Galactic latitude HI cirrus Ursa Major have similar 3D spectral index and that their spatial fluctuations share similar structures, despite showing a moderate correlation. These results support the fact that for statistical measurements over a region, the amount of velocity fluctuations as a function of scales has an impact on the gas density fluctuations, even if velocity and density fluctuations are not expected to be perfectly correlated.

\begin{figure}
\centering
\includegraphics[width=0.5\textwidth]{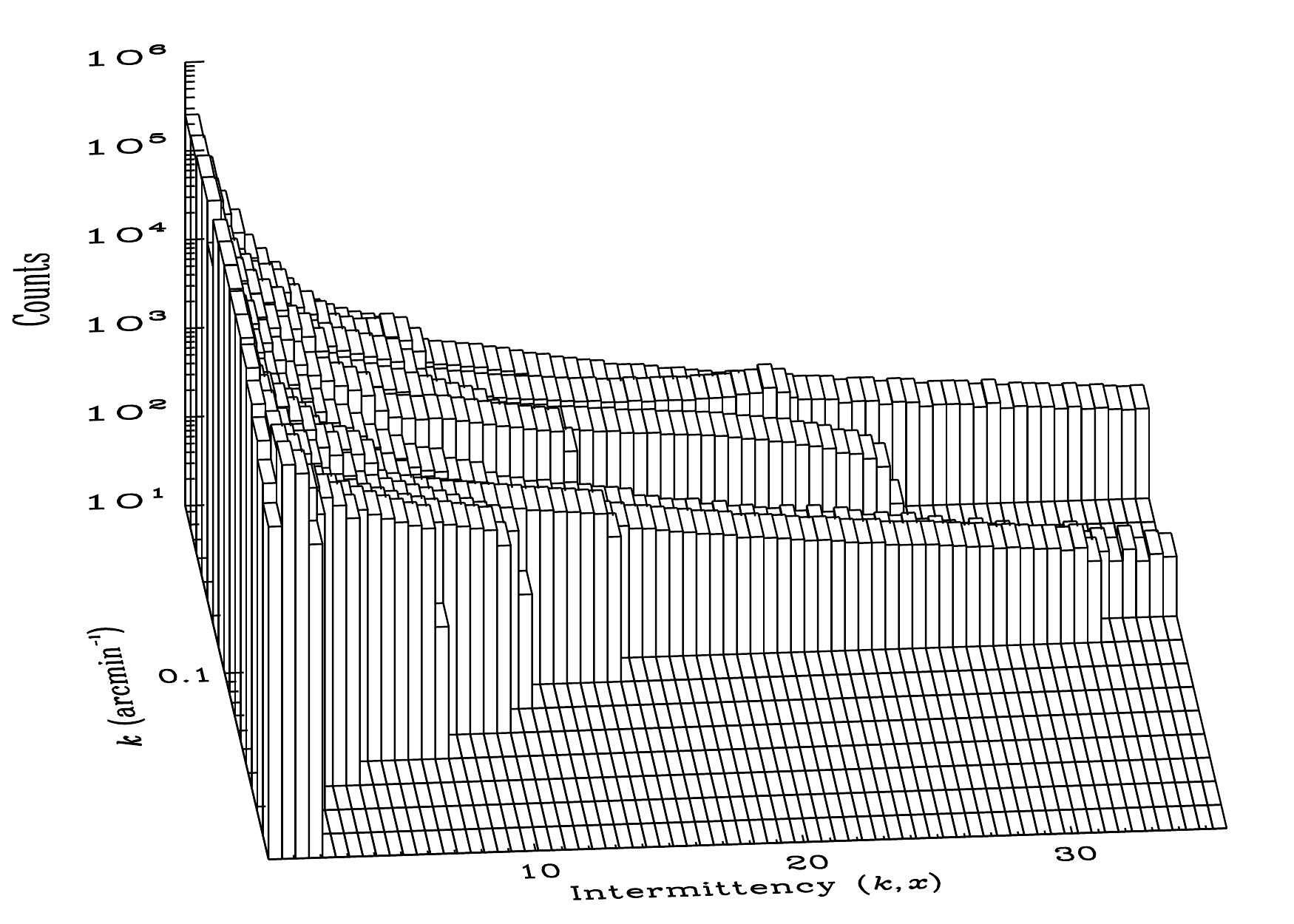}
\caption{Plots of the intermittency measure $I(l,\vec{x})$, as a function of the spatial frequency $k$. Frequencies from $\sim 0.01$ to $\sim 0.75$ arcmin$^{-1}$ are plotted, which approximately correspond to scales where enough statistically independent points are available at large scales and where the signal is not dominated by the noise at small scales.}
\label{fig:PDF_power}
\end{figure}

In Fig. \ref{fig:PDF_power}, we display the normalised PDFs of the squared amplitude of complex wavelet coefficients as a function of spatial scales. This plot corresponds to the intermittency measure as defined by \citetads{1992AnRFM..24..395F}:

\begin{equation}
I(l,\vec{x}) = \frac{|\tilde{f}(l,\vec{x})|^2}{\langle |\tilde{f}(l,\vec{x})|^2 \rangle_{\vec{x}}}
\label{eq:intermittency}
\end{equation}

\begin{figure}
\centering
\includegraphics[width=0.5\textwidth]{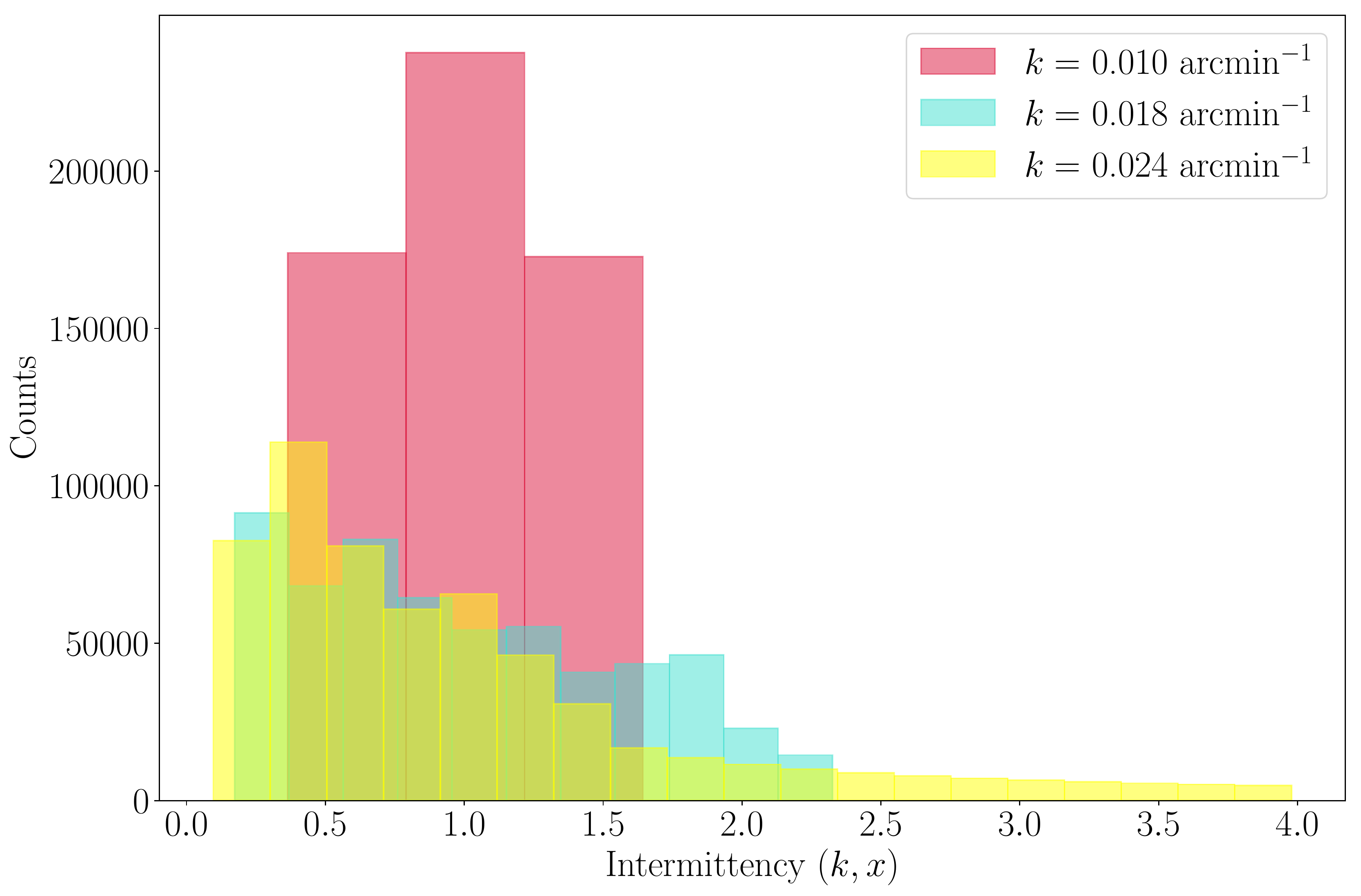}
\caption{The intermittency measure in linear scale for three spatial frequencies. Compared to Fig. \ref{fig:PDF_power}, the number of bins at each spatial scale respects the relation $n=(N_x\times N_y)/l^2$, where $k=|\vec{k}_0|/l$.}
\label{fig:PDF_power_solo}
\end{figure}

\noindent As described by \citetads{1992AnRFM..24..395F}, if $I(l,\vec{x}) = 1$ for all $\vec{x}$ and for all $l$, then it means that there is no flow intermittency. In that case, each location $\vec{x}$ would have the same power spectrum, which corresponds to what we expect from a Fourier power spectrum. Figure \ref{fig:PDF_power} shows extreme departures from the mean power value at almost every spatial scales above 0.025 arcmin$^{-1}$. The intermittency measure shows that many locations in the maps contribute more than 30 times than the average over $\vec{x}$ to the Fourier power spectrum for a broad range of scales. The intermittency PDFs in Fig. \ref{fig:PDF_power} are calculated for a constant bin of 0.5, however the number of pixels statistically independent varied as a function of spatial scales. The number of independent pixel is determined following the relation $n=(N_x\times N_y)/l^2$. Figure \ref{fig:PDF_power_solo} shows the intermittency measure in linear scale for three spatial frequencies smaller than 0.1 arcmin$^{-1}$, where the number of independent pixel is respected.

\begin{figure}
\centering
\includegraphics[width=0.5\textwidth]{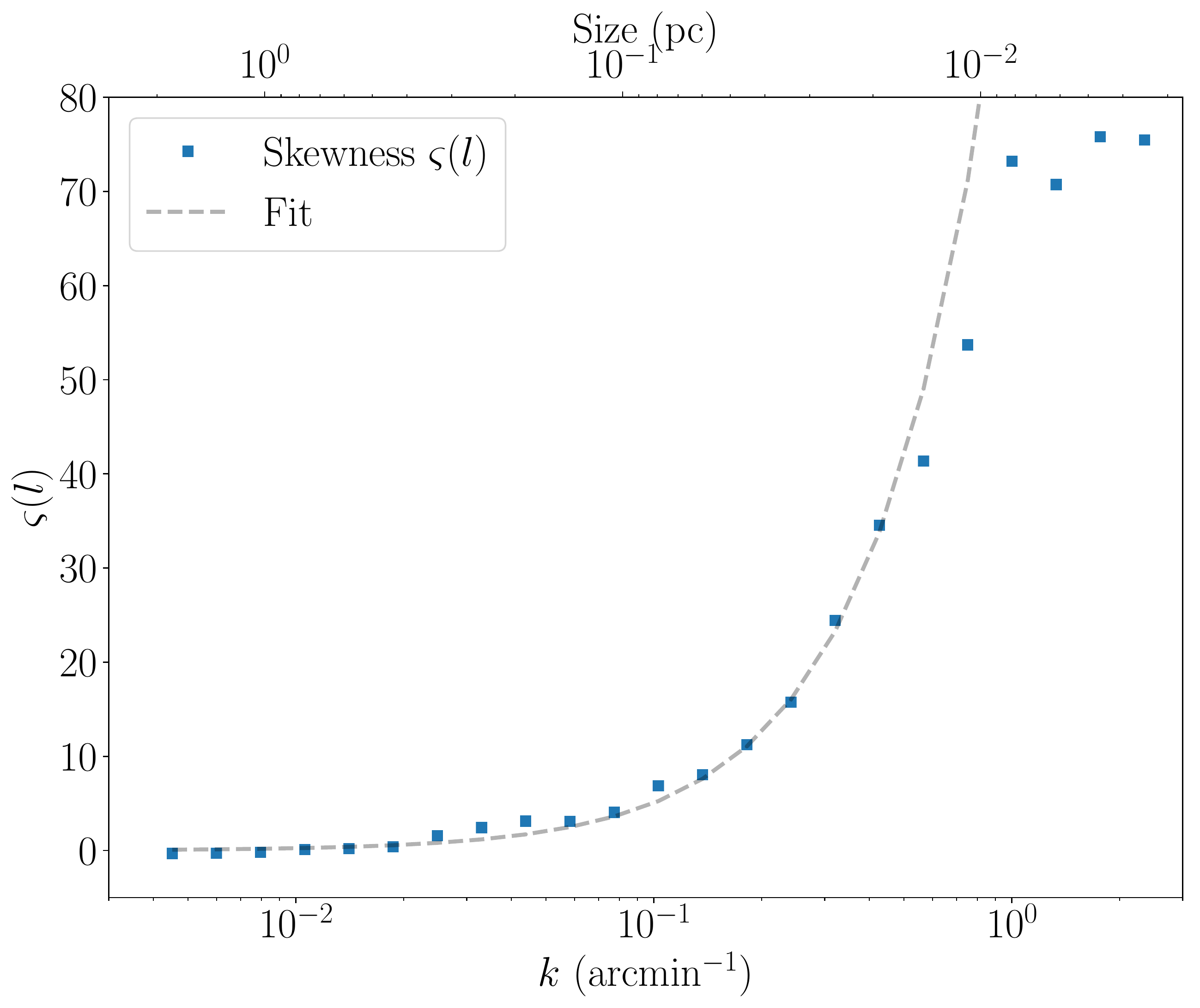}
\caption{The skewness, $\varsigma(l)$, for the intermittency measures defined in equation \ref{eq:intermittency} and shown Fig. \ref{fig:PDF_power} and \ref{fig:PDF_power_solo}}
\label{fig:PDF_skewness}
\end{figure}

The skewness of the intermittency PDFs has been calculated as a statistical measurement of the increasing intermittency as a function of the spatial scale. The skewness is defined as:

\begin{equation}
\varsigma(l) = \frac{\langle (I(l,\vec{x}) - \langle I(l,\vec{x}) \rangle_{\vec{x}})^3 \rangle_{\vec{x}}}{\sigma_l^3},
\label{eq:skewness}
\end{equation}

\noindent where $\sigma_l$ is the standard deviation of the intermittency measure $I(l,\vec{x})$ for the given scale $l$. Figure \ref{fig:PDF_skewness} shows the skewness value, $\varsigma$, for each spatial scale converted in wavenumber $k$. It can be seen that $\varsigma(k)$ increases exponentially towards smaller scales until scales become dominated by the noise level. The skewness value follows the fitted relation, $\varsigma(k) = (103\pm1) \times k^{1.31 \pm 0.08}$. A small deviation from the exponential fit is present at $0.02 \lesssim k \lesssim 0.08$ arcmin$^{-1}$, which corresponds to $0.1 \lesssim l \lesssim 0.3$ pc. This deviation will be discussed furthermore in section \ref{sec:ratio}.

\subsection{Multiscale non-Gaussian segmentation}

\begin{figure}
\centering
\includegraphics[width=0.48\textwidth]{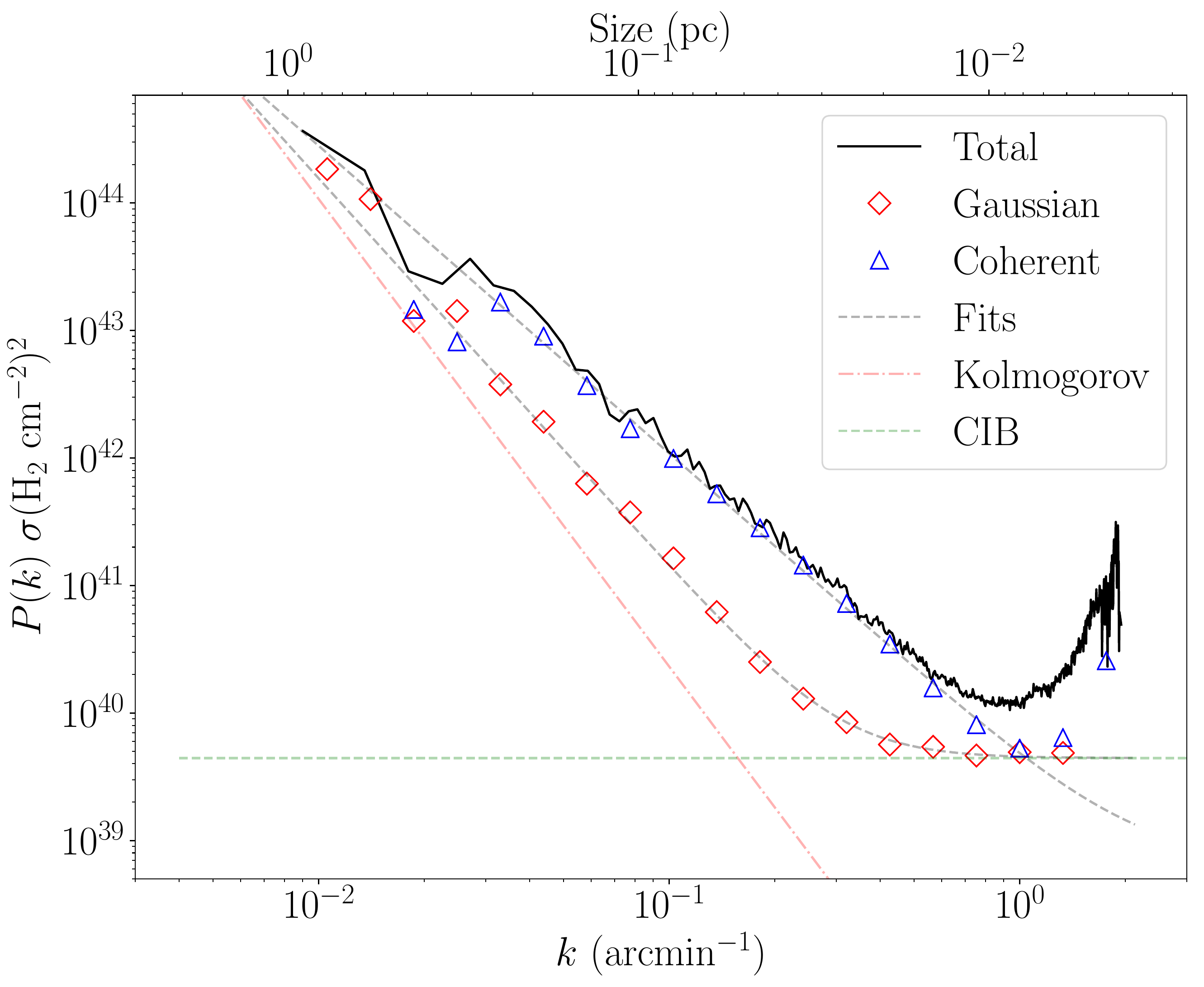}
\caption{Segmented power spectra for the Polaris flare column density map. The total Fourier power spectrum shown in Fig. \ref{fig:polaris_pow_spec} is represented by the solid black line. The red diamonds show the power spectrum for the Gaussian self-similar part of the map. The blue triangles show the power spectrum for the non-Gaussian coherent part of the map. The black dashed line represents the fits of the curves and the red dotted line represents the CIB and sources contribution level.}
\label{fig:polaris_seg_pow_spec}
\end{figure}

This section presents the result of the MnGSeg technique applied on the Polaris flare region. Power spectra of both components are shown in Fig. \ref{fig:polaris_seg_pow_spec}, the Gaussian and the coherent parts, where the noise level is subtracted and the telescope transfer function is divided. As seen in the intermittency PDFs of wavelet coefficients in the previous section, intermittency in density starts to appear at large spatial scale leading to the segmentation of coherent features by the MnGSeg technique from $k \gtrsim 0.02$. It corresponds to a spatial scale of $\sim 0.4$ pc. The coherent features are then dominating at all lower scales and follow closely the total power law measured in the Fourier power spectrum. The Gaussian part of the segmentation has also a power law followed by a complete flattening at small scales. According to the previous model of the Fourier power spectrum described in equations \ref{eq:MAMD_model} and \ref{eq:MAMD_Psky}, it is now possible to propose a more detailed model taking into account respectively the Gaussian and coherent power spectra, $P^G(k)$ and $P^C(k)$:

\begin{equation}
P(k) = P^G(k) + P^C(k),
\label{eq:sky_model}
\end{equation}

\noindent where

\begin{equation}
P^G(k) = \phi(k) ( A_{\textrm{ISM}}^G k^{-\gamma_G} + P_{\textrm{CIB}}(k) ) + N(k)
\label{eq:gaussian_model}
\end{equation}

\noindent and

\begin{equation}
P^C(k) = \phi(k)( A_{\textrm{ISM}}^C k^{-\gamma_C} + P_{\textrm {src}}).
\label{eq:coherent_model}
\end{equation}

\noindent Since the noise component of the signal usually respects a distribution close to a normal distribution, it should now be present entirely in the Gaussian segmented part. This argument should also be generally true for the CIB component, except for the brightest and more nearby galaxies which are found in the small-scale non-Gaussian component. The CIB signature can be easily associated with the flatten part of the Gaussian power spectrum where coherent structures, at the same spatial scales, are still contributing to the coherent power law, i.e. for $0.3 \leq k \leq 1.0$ arcmin$^{-1}$. A comparison with the CIB component extracted from the Polaris region is done at 250 $\mu$m in the next section. The values for the power spectra fits, according to equations \ref{eq:gaussian_model} and \ref{eq:coherent_model} are summarised in Table \ref{tab:pow_spect_fits_cohe_incohe}. Both wavelet power spectra were fitted between $0.03 \leq k \leq 1.0$ arcmin$^{-1}$ after being subtracted by the noise level, for the Gaussian part only, and divided by the telescope transfer function for both spectra. The coherent power law fit corresponds, within the uncertainties, to the power law estimated for the total Fourier power spectrum. On the other hand, the Gaussian power law fit is steeper and closer to the Kolmogorov power law of $-11/3$ for a three-dimensional turbulent medium.

\begin{table}
\centering
\caption{Fit values for the column density map Gaussian and the non-Gaussian power spectra}
\label{tab:pow_spect_fits_cohe_incohe} 
\resizebox{\linewidth}{!}{%
\begin{tabular}{lccc}
\hline\hline
&$A_{\textrm{ISM}}$ & Power-Law $(\gamma)$ & $P_{\rm{CIB/sources}}$ \\
&(H$_2$ cm$^{-2}$)$^2$ & & (H$_2$ cm$^{-2}$)$^2$ \\ \hline

Gaussian & $(1.3\pm0.2) \times 10^{38}$ & $3.05\pm0.07$ & $(4.4\pm0.3) \times 10^{39}$  \\
Coherent & $(4.2\pm0.4) \times 10^{39}$ &  $2.41\pm0.04$ &$(7\pm8)\times 10^{38}$  \\

\hline
\end{tabular}}
\end{table}

\begin{figure*}
\centering
\includegraphics[width=1.0\textwidth]{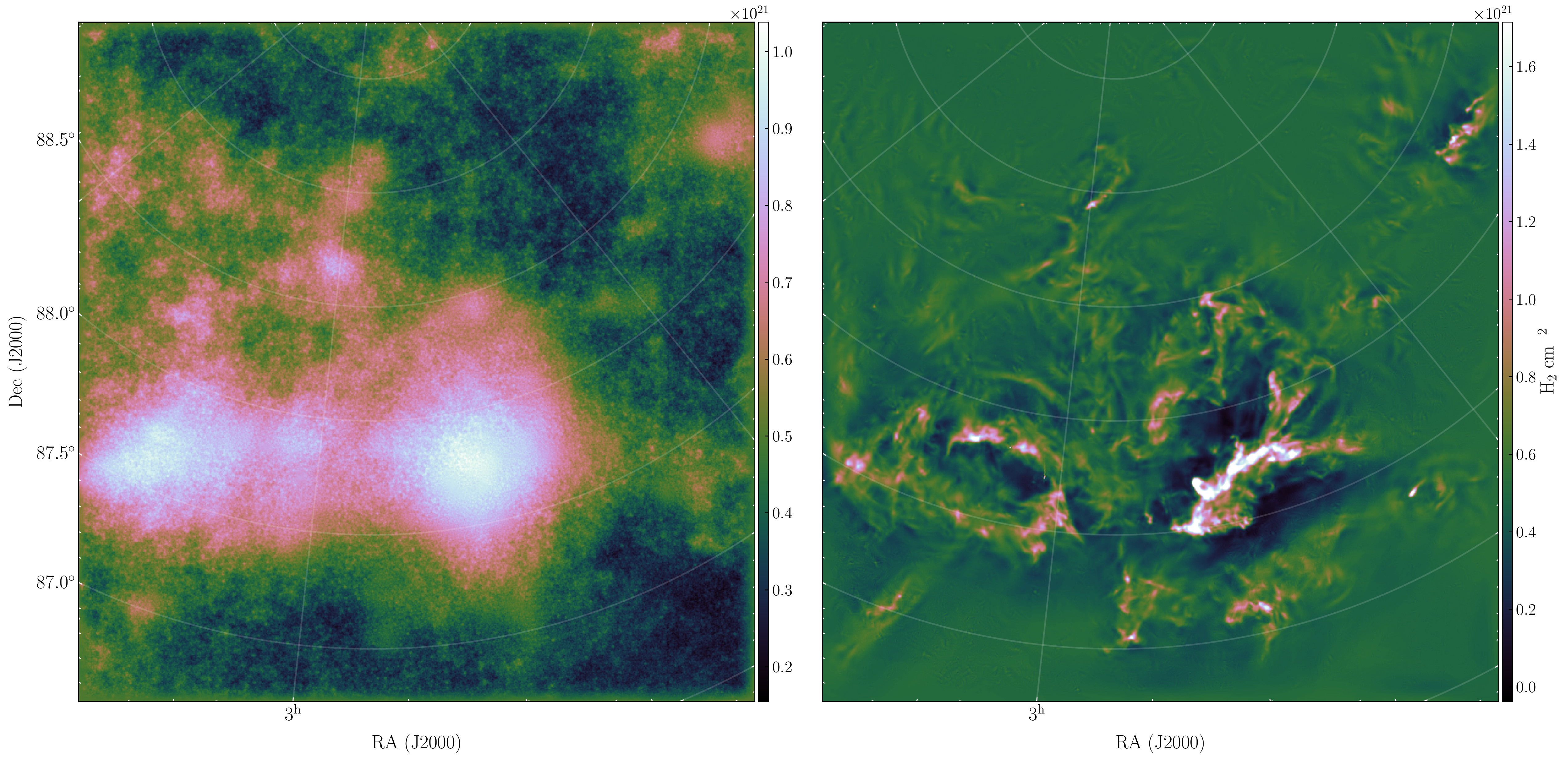}
\caption{The Gaussian (left) and coherent (right) reconstructed column density maps following equations \ref{eq:synthesisG} and \ref{eq:synthesisC}. For these two reconstructions, the correction factor, $C_{\delta}$, has been multiplied to both maps, and the mean values of the initial map, $\mu_0$, has been distributed to both maps.}
\label{fig:reconstruction}
\end{figure*}

\begin{figure}
\centering
\includegraphics[width=0.48\textwidth]{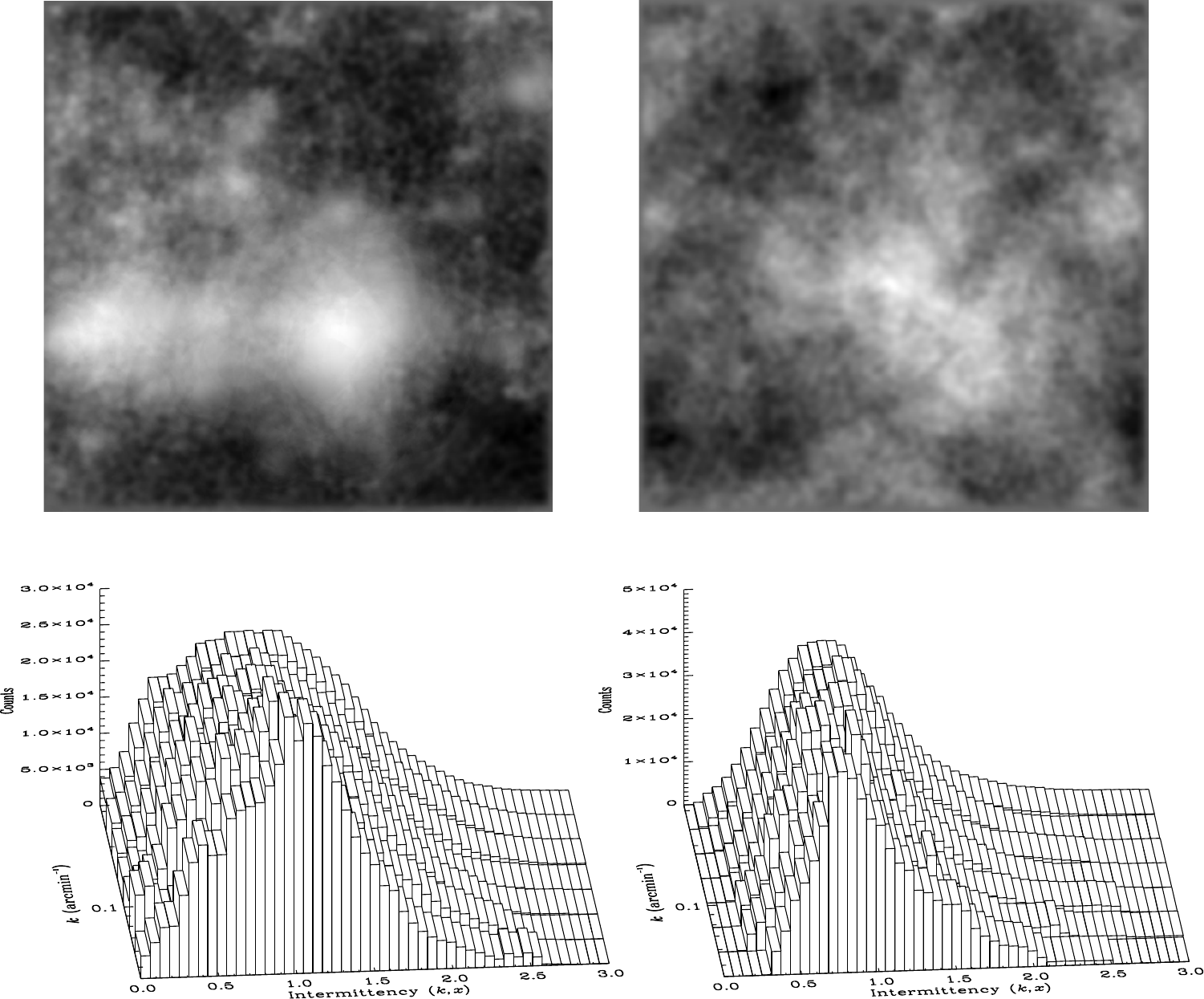}
\caption{Top left panel: the Gaussian reconstructed map filtered for scales $k \gtrsim 0.3$ arcmin$^{-1}$. Top right panel: fBm simulation with the same power law as the Gaussian part of Polaris. The fBm is also filtered for $k \gtrsim 0.3$ arcmin$^{-1}$ Bottom left panel: the intermittency measure as defined in eq. \ref{eq:intermittency} for the Gaussian segmentation of wavelet coefficients. Bottom right panel: the intermittency measure of the fBm simulation for the same spatial scales than Polaris.}
\label{fig:polaris_fbm_pdfs}
\end{figure}

The Gaussian and coherent reconstructed maps following the relations \ref{eq:synthesisG} and \ref{eq:synthesisC} are shown in Fig. \ref{fig:reconstruction}. The Gaussian map shows smooth features dominated by large-scale density fluctuations, as it is measured by the steeper power law of its power spectrum. The small-scale fluctuations are dominated by a granular component characterised by the flat part of the Gaussian power spectrum and dominated by the CIB signal. The PDFs associated to the Gaussian map are compared to a fBm in Fig. \ref{fig:polaris_fbm_pdfs}. Both maps are filtered for $k \gtrsim 0.3$ arcmin$^{-1}$ in order to filter the CIB and the instrumental noise in the case of the Polaris map. The Gaussian PDFs show that the segmentation algorithm successfully removed most of the intermittent coefficients compared to the original intermittency measure in Fig. \ref{fig:PDF_power}. After the segmentation, most of the intermittency measure is contained between $0 < I(l,\vec{x}) \lesssim 2.5$ and, in contrast with the original intermittency measure, all the distribution are now centred on 1.0 and are similar across the wavenumber $k$. Compared to the fBm simulation, the PDFs associated with the Gaussian density fluctuations of Polaris are broader. This may come from the fact that the Polaris region shows more areas of low density than the fBm. In fractal analysis, this aspect referred to the \emph{lacunarity} of the structures \citepads{1982fgn..book.....M}. Even if two fractal images share the same power law, their appearance can differ according to their different distribution $I(l,\vec{x})$. In this case the main difference is that the fBm image, constructed in the Fourier space, is made only of stationary fluctuations, which is not necessary the case for the ISM density fluctuations.

\begin{figure}
\centering
\includegraphics[width=0.48\textwidth]{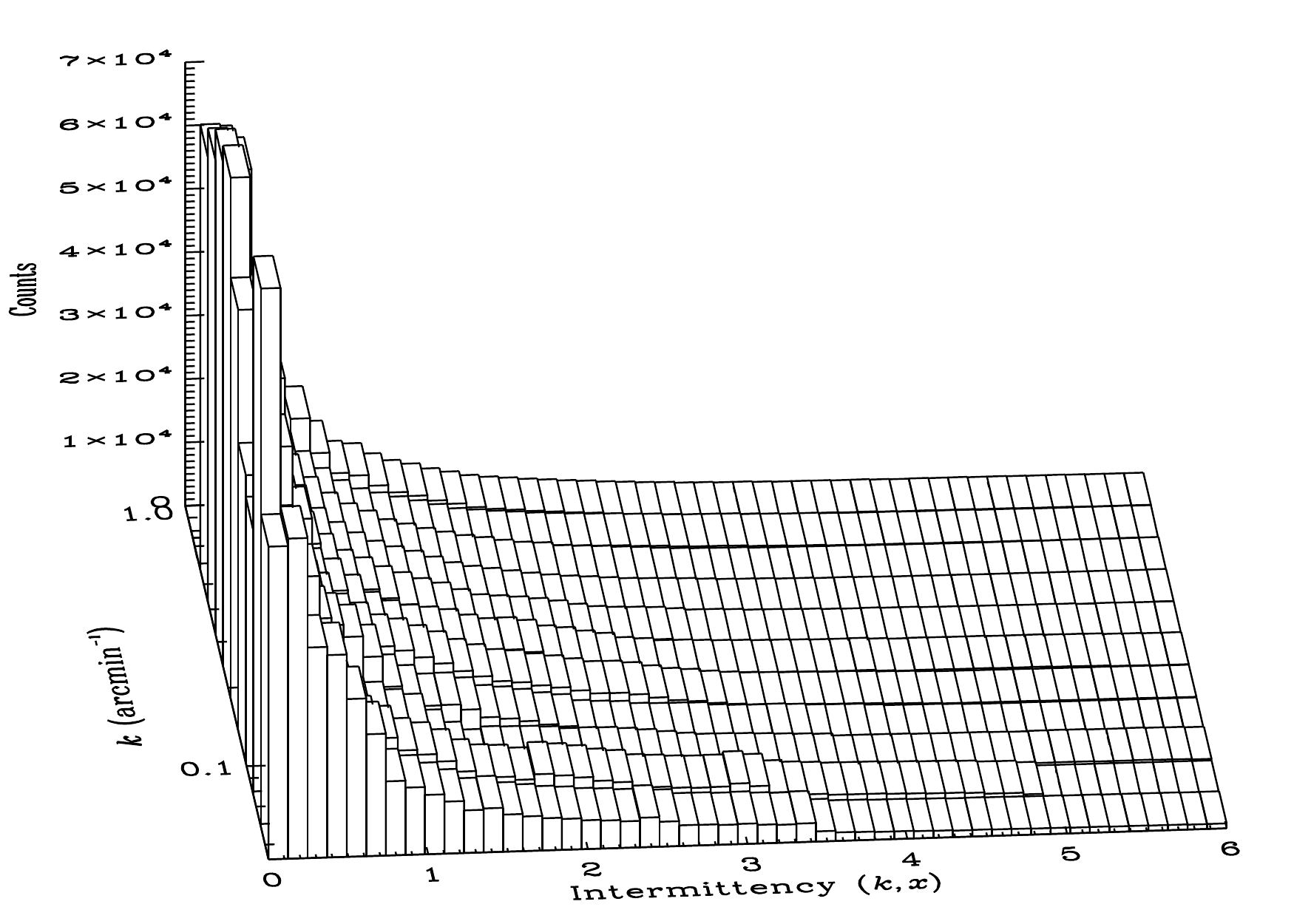}
\caption{The intermittency measure as defined in eq. \ref{eq:intermittency} for the non-Gaussian wavelet coefficients.}
\label{fig:polaris_coherent_pdfs}
\end{figure}

The coherent map in Fig. \ref{fig:reconstruction} is dominated by the denser elongated structures. According to its power spectrum in Fig. \ref{fig:polaris_seg_pow_spec}, the non-Gaussian fluctuations are present on a broad range of spatial scales. A closer inspection of the coherent map indeed indicates that most of the filamentary structures are embedded in larger structures that have been identified as non-Gaussianities. No particular break is visible on the coherent power spectrum, which could be interpreted as no spatial scales being predominant for the non-Gaussianities. This result is in agreement with the recent analysis of Polaris filamentary structures done by \citetads{2019A&A...621A...5O}, where the Polaris Flare shows an almost scale-free filamentary spectrum. However, as defined in equations \ref{eq:Fan_wavelet} and \ref{eq:global_scalogram}, the power spectrum is only sensitive to the mean value of the power distribution. As it can be seen in Fig. \ref{fig:polaris_coherent_pdfs}, the intermittency measure for the non-Gaussian wavelet coefficients is largely skewed and not centred on $I(k,\vec{x})=1.0$, which means that the coherent wavelet power spectrum, as it is the case for the Fourier power spectrum, is not directly representative of the underlying power distribution. However, the unique power law associated with the non-Gaussianities is still a very interesting result and can be an indication that the non-Gaussianities as a function of scales are linked to the inertial range of the Gaussian part. This result might possibly be a direct measurement of the inertial-range intermittency detected on a density map of a star-forming region. This hypothesis is discussed further in section \ref{sec:discussion}.

\subsection{CIB measurement \& extraction}\label{sec:CIB}

The CIB corresponds to high-redshift starburst galaxies that are unresolved by far-infrared and submillimeter observations \citepads{1996A&A...308L...5P, 1998ApJ...508..123F, 1999A&A...344..322L}. The energy peak of this signal is around 200 $\mu$m. Because the redshift distribution of these dusty star-forming galaxies is relatively broad, the signal is also changing as a function of the observed wavelength \citepads{2012A&A...542A..58B}. For this reason, using the column density map is not appropriate to analyse the CIB signature we detected in the Polaris flare region. A second segmentation analysis has therefore been done on the Herschel 250 $\mu$m map alone. Accurate CIB measurements using \emph{IRAS}, \emph{Planck} and \emph{Herschel} observations have already been done \citepads{2011A&A...536A..18P, 2012A&A...543A.123P, 2013ApJ...772...77V, 2013ApJ...779...32V}. Our analysis on the 250 $\mu$m map will be compared with \citetads{2013ApJ...772...77V} results on the Multi-tiered Extragalactic Survey \citep[HerMES;][]{2012MNRAS.424.1614O}.

\begin{figure}
\centering
\includegraphics[width=0.48\textwidth]{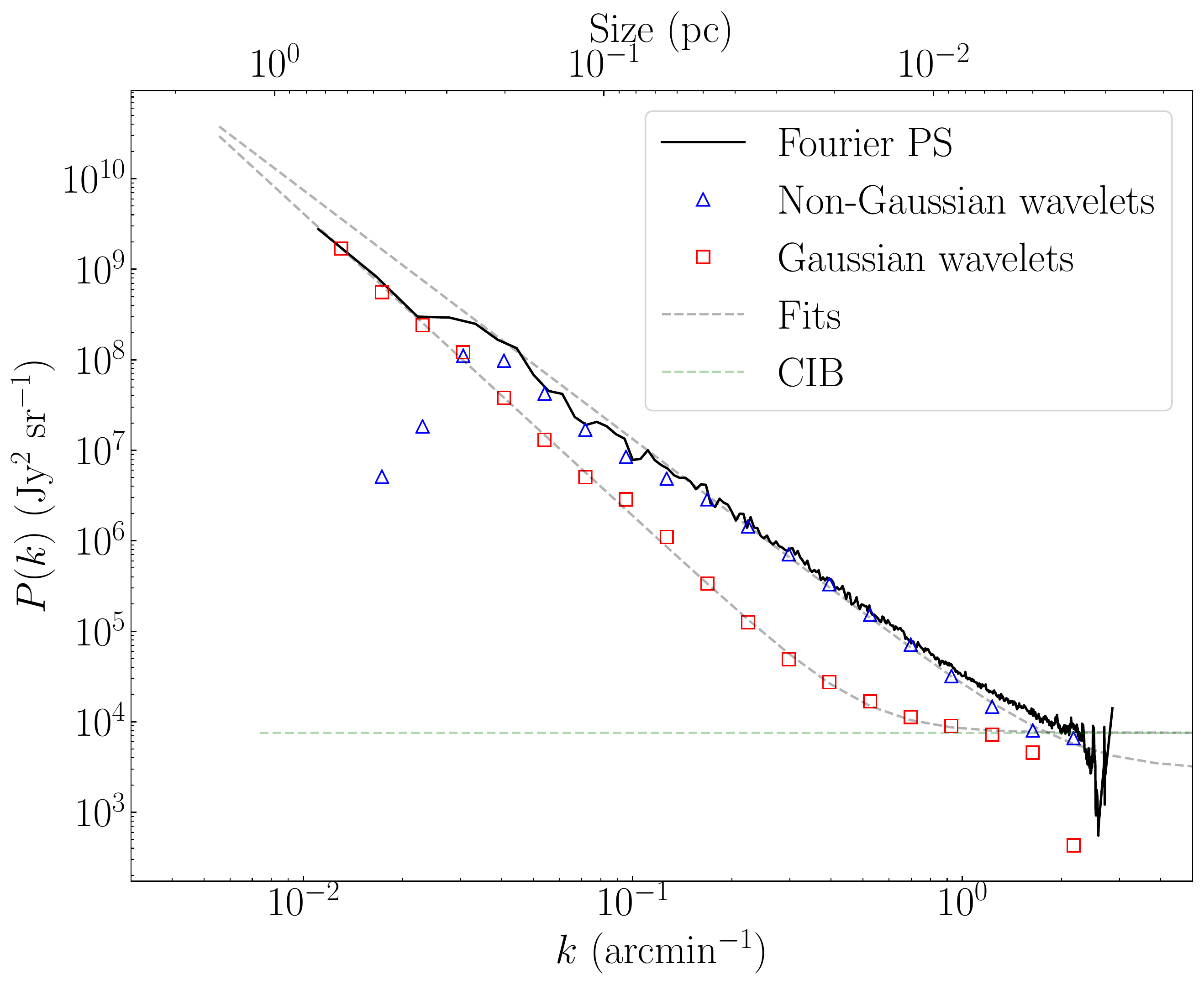}
\caption{Segmented power spectra for the Polaris flare map at 250 $\mu$m. The total Fourier power spectrum is represented by the solid black line. The red diamonds show the power spectrum for the Gaussian self-similar part of the map. The blue triangles show the power spectrum for the non-Gaussian coherent part of the map. The black dashed line represents the fits of the curves and the green dashed line represents the CIB and sources contribution level. Fitted values are listed in Table \ref{tab:pow_spect_fits_250}.}
\label{fig:polaris250_pow_spec_seg}
\end{figure}

The segmented (Gaussian and coherent) power spectra for the \emph{Herschel} 250 $\mu$m map are presented in Fig. \ref{fig:polaris250_pow_spec_seg}. The spectra have been calculated in the same way as for the column density map. The fitted values are listed in Table \ref{tab:pow_spect_fits_250}. The general shapes of the spectra are similar to the ones calculated for the column density map in Fig. \ref{fig:polaris_seg_pow_spec}. The fitted values for the total Fourier power spectrum corresponds within uncertainties with the values fitted by \citetads{2010A&A...518L.104M} for the Herschel 250 $\mu$m map of Polaris covering a larger and slightly different field of view. Contrary to the column density map, the power for the Gaussian power spectrum is dropping quickly for $k  \gtrsim 1$ arcmin$^{-1}$. This difference can be attributed to the smaller beam pixel size at 250 $\mu$m.

\begin{table}
\centering
\caption{Fit values for the total 250 $\mu$m map, the Gaussian and the non-Gaussian power spectra}
\label{tab:pow_spect_fits_250} 
\resizebox{\linewidth}{!}{%
\begin{tabular}{lccc}
\hline\hline
&$A_{\textrm{ISM}}$ & Power-Law $(\gamma)$ & $P_{\rm{CIB/sources}}$ \\
&(Jy$^2$sr$^{-1}$) & & (Jy$^2$sr$^{-1}$) \\ \hline

Fourier     & $(3.07\pm0.03) \times 10^{4}$ & $2.58\pm 0.01$ & $(3.6\pm0.1) \times 10^{3}$\\
Gaussian & $(8\pm1) \times 10^{2}$ & $3.34\pm0.05$ & $(7.6\pm0.7) \times 10^{3}$  \\
Coherent & $(2.3\pm0.2) \times 10^{4}$ &  $2.75\pm0.07$ &$(2.9\pm0.8)\times 10^{3}$  \\

\hline
\end{tabular}}
\end{table}

\citetads{2013ApJ...772...77V} measured from the combination of different Herschel fields and extended source masked at $k = 1.406$ arcmin$^{-1}$ a power of $(8.54 \pm 0.31) \times 10^3$ Jy$^2$ sr$^{-1}$. Their spectra are corrected for a cirrus power law fixed to $\gamma = 3.0$. Here, we considered the power law of the Gaussian part only as the cirrus signal. Following equation \ref{eq:gaussian_model}, we find a power of $(7.6 \pm 0.7) \times 10^3$ Jy$^2$ sr$^{-1}$. This value of the CIB power corresponds very well to the one evaluated previously by \citetads{2013ApJ...772...77V}. It is important to recall that in our case the CIB signal was  dominated by the foreground emission of Polaris. As we can see in Fig. \ref{fig:polaris250_pow_spec_seg}, the small-scale CIB power was dominated by the power of coherent structures and the MnGSeg method succeeded nonetheless to extract the mean power of this relatively faint signal over the map. A more in-depth analysis of the CIB signal is out of the scope for this paper, but the MnGSeg method presents itself as a good strategy for this analysis.

\begin{figure*}
\centering
\includegraphics[width=1.0\textwidth]{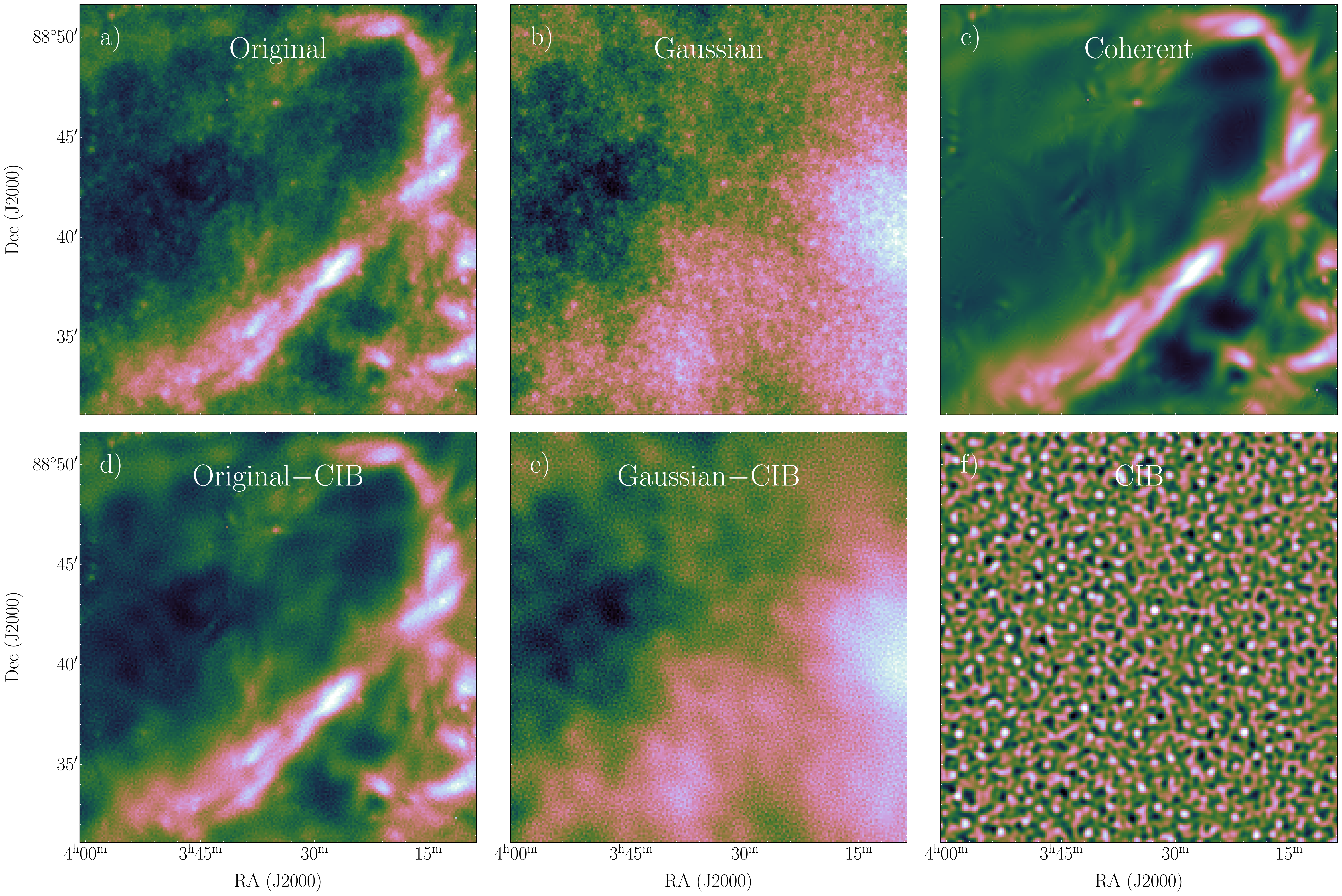}
\caption{Different segmentations applied on a subregion of the Polaris map at 250 $\mu$m. a)The original map. b) The Gaussian map reconstructed from all spatial scales. c) The coherent map reconstructed from all spatial frequencies. d) The original map without the frequencies associated with the flatten and dropping part of the Gaussian power spectrum related to the CIB, i.e. $0.9 \lesssim k \lesssim 2.2$ arcmin$^{-1}$. e) The Gaussian map without frequencies $0.9 \lesssim k \lesssim 2.2$ arcmin$^{-1}$. f) Summation of the Gaussian frequencies $0.9 \lesssim k \lesssim 2.2$ arcmin$^{-1}$ dominated by the CIB signal.}
\label{fig:Polaris250_CIB_zoom}
\end{figure*}

\subsection{Ratio coherent/Gaussian}\label{sec:ratio}

\begin{figure}
\centering
\includegraphics[width=0.49\textwidth]{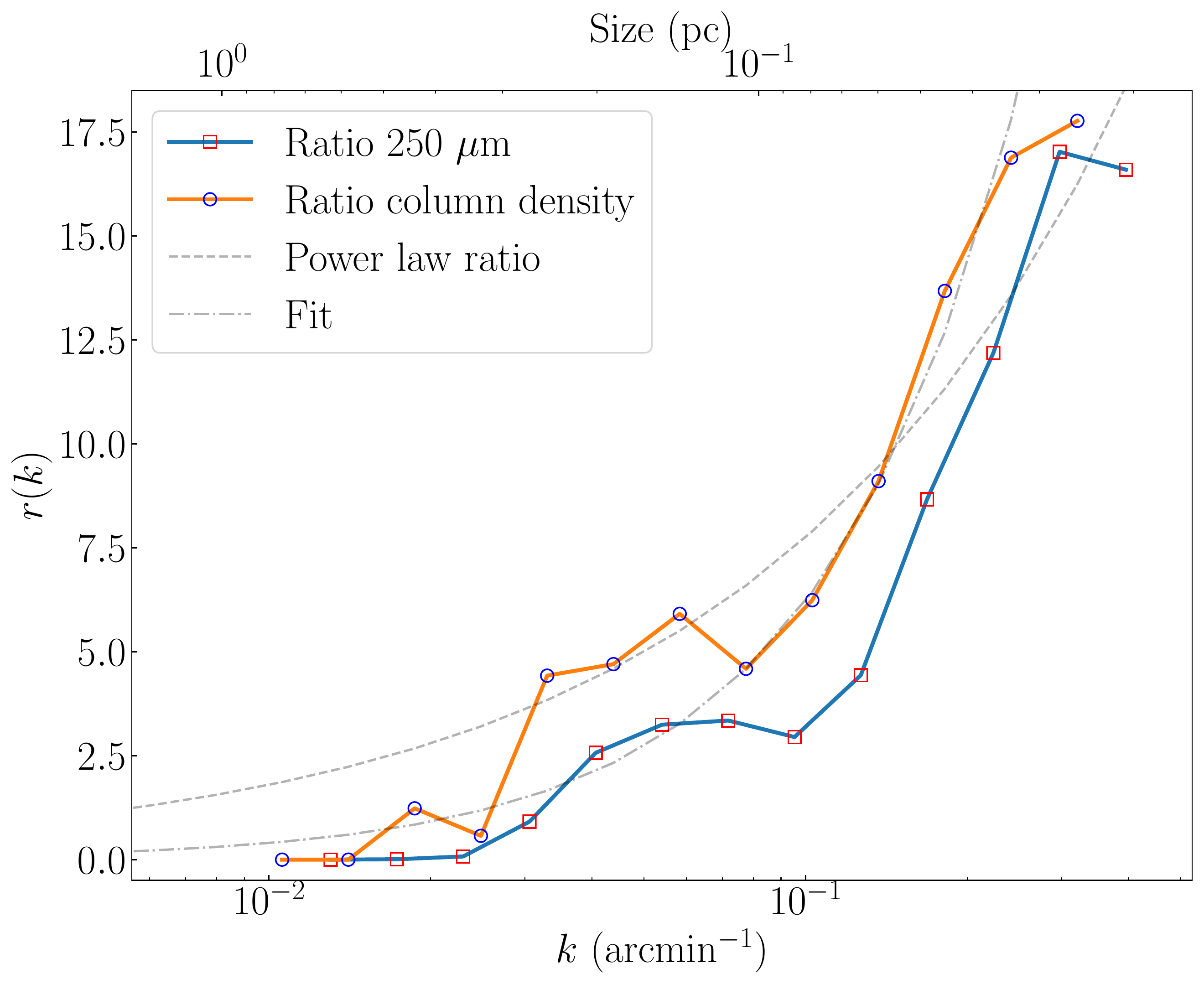}
\caption{The ratio between the coherent and Gaussian power spectra for the column density and the 250 $\mu$m maps. The dashed curve shows the exponential curve as defined in equation \ref{eq:ratio} using the fitted power law values for the column density map in Table \ref{tab:pow_spect_fits_cohe_incohe}. The dot-dashed curve shows the exponential for the power spectra ratio fitted for $0.08 \lesssim k \lesssim 0.2$ arcmin$^{-1}$, which corresponds to $0.03 \lesssim l \lesssim 0.1$ pc.}
\label{fig:ratio}
\end{figure}

As shown in Fig. \ref{fig:PDF_skewness} and descfibed in section \ref{sec:intermittency}, the skewness of the intermittency PDFs increases exponentially towards smaller scales until scales become dominated by the noise level. Another way to look at the intermittency measure is to compare both components by calculating the ratio between the coherent and Gaussian power spectra. The ratios for the column density and the 250 $\mu$m maps are plotted in Fig. \ref{fig:ratio}. The ratios are  corrected for the noise and CIB/sources contributions and, according to equations \ref{eq:gaussian_model} and \ref{eq:coherent_model}, it is defined as,

\begin{equation}
r(k) = \frac{A^C}{A^G} k^{-\gamma_C+\gamma_G}.
\label{eq:ratio}
\end{equation}

For both maps, the power spectra ratio have a power law shape as defined in equation \ref{eq:ratio} with a bump centred at $k\approx0.05$ arcmin$^{-1}$, which corresponds to $\sim 0.15$ pc. This bump was also noticed in Fig. \ref{fig:PDF_skewness} for the skewness as a function of scales. The dashed curve in Fig. \ref{fig:ratio} shows the exponential curve as defined in equation \ref{eq:ratio} using the fitted power law values for the column density map in Table \ref{tab:pow_spect_fits_cohe_incohe}. The dot-dashed curve shows the exponential for the power spectra ratio fitted for $0.077 \lesssim k \lesssim 0.24$ arcmin$^{-1}$ and has a power law of $1.2\pm0.1$. Assuming that the power law fit showed in Fig. \ref{fig:polaris_seg_pow_spec} was also affected by this bump around $k\approx0.05$ arcmin$^{-1}$, the corrected Gaussian power law would become $\gamma_r + \gamma_C= 3.6 \pm 0.1$, which corresponds to the Kolmogorov power law for a 3D incompressible turbulent medium.

\begin{figure*}
\centering
\includegraphics[width=1.0\textwidth]{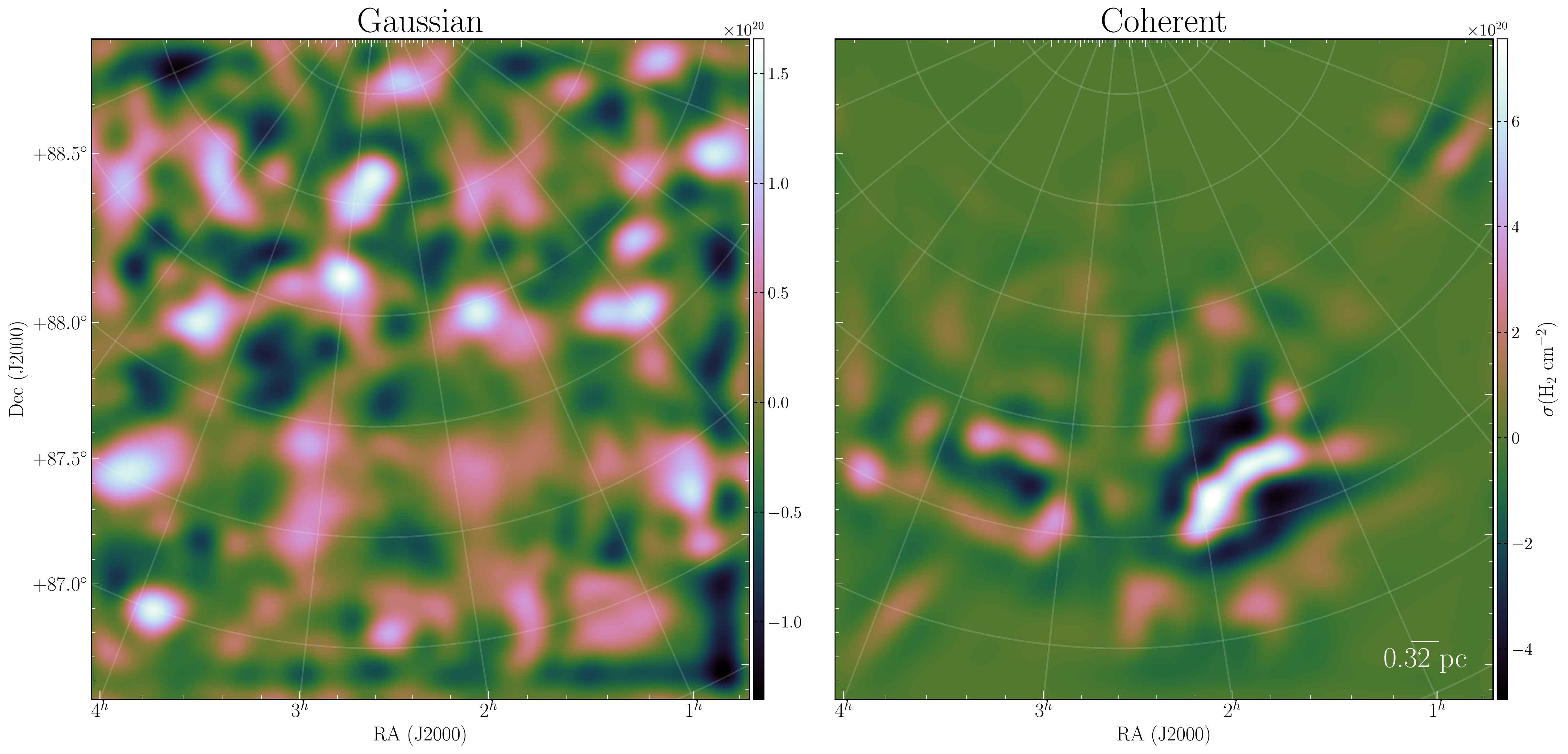}
\caption{The Gaussian and coherent reconstructed maps for $0.025 \lesssim k \lesssim 0.077$ arcmin$^{-1}$, which corresponds to $0.11 \lesssim l \lesssim 0.33$ pc.}
\label{fig:bump}
\end{figure*}

The fact that this excess of power is present in both the column density map and the 250 $\mu$m map confirms that the excess is real and not an artefact from the column density calculation. Figure \ref{fig:bump} shows the Gaussian and coherent reconstructed maps for $0.025 \lesssim k \lesssim 0.077$ arcmin$^{-1}$, which corresponds to $0.11 \lesssim l \lesssim 0.33$ pc. As expected, the Gaussian map shows more uniform density fluctuations than the coherent map. For this range of spatial scales, the excess of power can be attributed to contrasted structures that can also easily be identified in Fig. \ref{fig:polaris}, as the ``saxophone'' in the south-west region of the field. However, the excess of power in the power spectra ratio is also present when the ``saxophone'', the brightest high column density region, is excluded. This means that more quiescent regions of the Polaris Flare, as the one defined by \citetads{2013ApJ...766L..17S}, also has exhibit this characteristic scale.

\subsection{Anisotropies}

\begin{figure}
\centering
\includegraphics[width=0.50\textwidth]{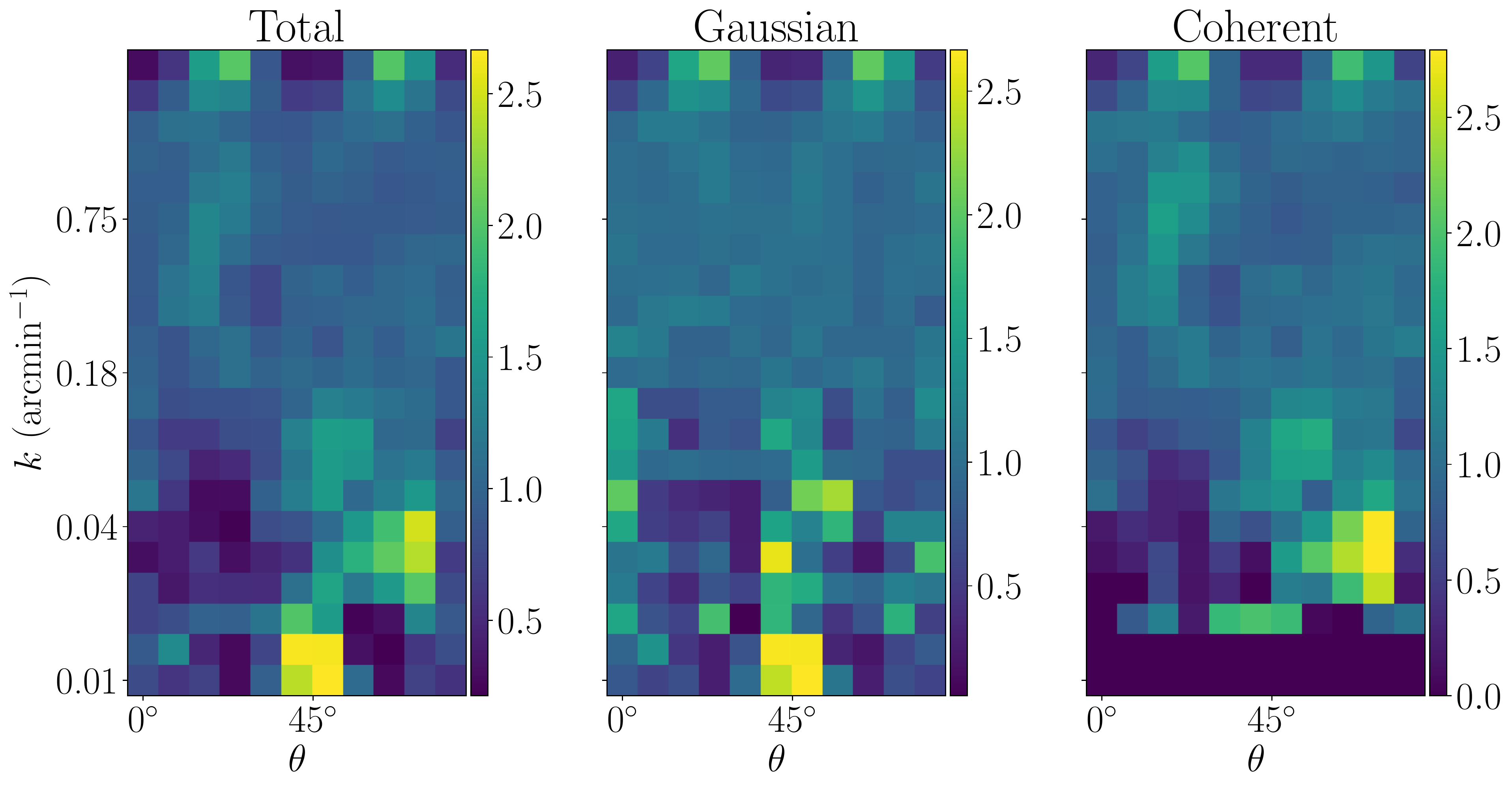}
\caption{The anisotropy measure as a function of the azimuthal angle $\theta$ and the wavenumber $k$ as defined by eq. \ref{eq:anisotropies}.}
\label{fig:anisotropies}
\end{figure}

Since the non-Gaussian segmentation is applied as a function of azimuthal angles, the segmentation should also reveal structure anisotropies as a function of spatial scales. Figure \ref{fig:anisotropies} shows the anisotropy measure as suggested by \citetads{1992AnRFM..24..395F} for the total set of wavelet coefficients, the Gaussian part and the coherent part. The anisotropy measure is defined as

\begin{equation}
A(l,\theta) = \frac{\sum_{\vec{x}}|\tilde{f}(l,\vec{x},\theta)|^2}{\langle \sum_{\vec{x}} |\tilde{f}(l,\vec{x},\theta)|^2 \rangle_{\vec{\theta}}}.
\label{eq:anisotropies}
\end{equation}

\noindent According to Fig. \ref{fig:anisotropies}, non-Gaussianities starts to appear in most directions at $k\approx0.018$ arcmin$^{-1}$. Larger anisotropies are measured at large scales around $45^{\circ}$ in the total set of coefficients and the Gaussian part whereas they are not detected as non-Gaussianities. It is important to recall that, at these large scales, the number of independent pixels is $n \approx 3$, thus the statistic is very low in order to detect any outlier. The next anisotropies are measured at $k\approx0.033$ arcmin$^{-1}$, where $n \gtrsim 20$ both in the Gaussian and coherent part of the field with a value of $A(l,\theta) \approx 2.5$, which is much lower than the intermittency measures shown in Fig. \ref{fig:PDF_power}. Anisotropies for $k\gtrsim0.06$ arcmin$^{-1}$ fluctuate slightly around 1.0, which means that, in the inertial range, structures have no particular direction. This result is also in agreement with the Polaris filamentary structure analysis of \citetads{2019A&A...621A...5O}. At the smallest scale, anisotropies are due to the pixelation effect. 

\section{Discussion}\label{sec:discussion}

The MnGSeg method succeed with simple assumptions, such as the self-similar nature of incompressible turbulence and the ergodicity inferred by the Fourier power spectrum analysis, to extract two fundamentally different statistical behaviours in molecular cloud gas density distributions. This dual nature of molecular clouds corresponds well to the previous description given by \citet{2004Ap&SS.292...89F}.  One fractal component, characterised by its self-similarity over the spatial scales and its unique power law, see Fig. \ref{fig:polaris_seg_pow_spec} and \ref{fig:polaris_fbm_pdfs}, corresponds to the diffuse component of the Polaris flare. The other component, called the coherent component, by opposition to the fractal one, displays a network of filaments of different sizes. These coherent filamentary structures are physically important, because they are where dense cores are embedded and thus crucial for star formation. \citetads{2010A&A...518L..92W}, identified the five densest cores within the MCLD 123.5+24.9 highest density structure (also called ``saxophone''). These starless cores have been subsequently studied for their chemical composition and their mass by \citetads{2012ApJ...745..195S} and \citetads{2015ApJ...809...17W}. Future analysis of the comparison between physical properties and spatial distribution of dense cores or young stellar objects and the coherent ISM structures will be important in order to shed the light of the impact of the ISM environment on the star formation processes.

The fact that filamentary structures affect the power spectrum analysis of a region was already noticed by \citetads{2011A&A...529A...1S} and \citetads{2014ApJ...788....3E}. However, in the present analysis, it is demonstrated that the fluctuations associated with the coherent filamentary component of the cloud are dominating over the major part of what is considered the inertial range of the power spectrum. This has been demonstrated in the present analysis on the Polaris flare and it has also been demonstrated previously on a larger area of the Galactic plane \citepads{2014MNRAS.440.2726R}. In the light of the present results, the important question to ask is not why filamentary structures do not produce any break in a power spectrum analysis, but how the coherent filamentary structures we extracted with the MnGSeg technique are related to the true scale-free component of molecular clouds. 

The segmentation procedure presented in this paper corresponds well to the bifractal intermittency model described on the review on turbulence by \citetads{1995turb.book.....F}. On a statistical and geometrical point of view, the Gaussian ``incompressible'' component extracted using the MnGSeg technique is monofractal by construction, which means that a single power law for the inertial range is sufficient to describe the dynamics of this component. In the bifractal model described by \citetads{1995turb.book.....F}, the inertial range of an intermittent turbulent flow is in fact the result of a competition between two power laws, where the one with the smallest exponent dominates at small scales. In the case where the steeper power corresponds to the Kolmogorov incompressible turbulence, only structure functions of order higher than 3 would be affected by the second power law. This means that the Fourier power spectrum is not sensitive to the bifractal model of intermittency. In our case, we show that the Gaussian part of the density map has a power law closer to the Kolmogorov incompressible turbulence, however the coherent part of the density map still has an influence on total Fourier power spectrum.

\citetads{1995turb.book.....F} also propose a more complete multifractal model of the intermittence. There is no reason to believe that the coherent component of the density map is monofractal. In fact, theories on fully developed turbulence predict that local variations of the turbulent energy cascade can be defined as a multifractal system \citepads{1995turb.book.....F, 1995PhyA..213..232A}. A full multifractal analysis of Polaris density and velocity field is beyond the scope of this paper. However, the extraction of a monofractal component covering all spatial scales and filling the whole field is a first step towards a better comprehension of how turbulence is related to the ISM gas structures. An important comparaison of multifractal analysis based on a box-counting approach has been recently done by \citetads{2018MNRAS.481..509E} between Hi-GAL observations, a key programme of \emph{Herschel}, fBm images and numerical simulations. They found that all the investigated fields, which are located in the Galactic plane, exhibit a multifractal rather than a simple monofractal structure. According to this result, numerical simulations appear, depending on the specific model, more similar to observations than fBms. Multifractal analysis methods of images or signals based on wavelet transforms are already developed and are actively used in multiple domains from turbulence analysis to medical research \citepads{2000EPJB...15..567A}, and the MnGSeg analysis could be potentially adapted to such approaches.

It is important to recall that the Kolmogorov energy cascade was derived from the velocity structure function and that our analysis is based on the density power spectrum. The link between the density and velocity power spectrum has been investigated previously mainly in the case of the warm ionised medium \citepads{1995ApJ...443..209A, 2007ApJ...665..402T}, where a power law of $\gamma = 11/3$ for the electron density power spectrum has been measured over a broad range of spatial scales. In hydrodynamics and magnetohydrodynamics, equations show that fluctuations in turbulence velocity, magnetic field and density are coupled. For subsonic to transonic turbulence, the velocity power spectrum hardly depends on the Mach number \citepads{2007ApJ...665..416K}. At higher Mach number, density fluctuations are more sensitive mainly due to the apparition of the compressive component of the turbulence, which is measured by the shallower energy power spectrum. The correlation between the density and velocity power spectra has also been investigated through the velocity-channel analysis \citep[VCA;][]{2000ApJ...537..720L}, which is based on the variations of the power spectra in velocity channels at changing velocity resolution.

As mentioned in section \ref{sec:assumptions}, the scale-free nature of ISM power spectra is commonly referred in the literature to the energy cascade of the Kolmogorov theory of turbulence, which predict a power law of $\gamma = 11/3$ for a subsonic and incompressible flow. This value is rarely found in the ISM and the typical value is closer to what is measured for the total Fourier power spectrum of the Polaris 250 $\mu$m map, i.e. $\gamma \approx 2.6$ \citepads{1997ApJ...474..730P}. The shallower slope is usually attributed to the presence of small scale structures present in the compressible ISM. Our analysis has demonstrated that the non-Gaussianities associated with the coherent structures in the Polaris molecular cloud are present on a wide range of spatial scales and not only at small scales. These non-Gaussianities appear progressively toward the small scales along the ``inertial range'' of the fractal diffuse component of the cloud. The Fourier power spectrum being equivalent only to the second order structure function, it is essentially not sensitive to these ``excess of power'' as a function of spatial scales, referring here to the non-Gaussianities shown in the intermittence measure of Fig. \ref{fig:PDF_power} and \ref{fig:PDF_power_solo}. The local reconstruction of the fluctuations responsible for these ``excess of power'' as a function of spatial scales shows that the non-Gaussianities are associated with the filamentary network of molecular clouds, see right panel of Fig. \ref{fig:reconstruction}. Following our algorithm, the non-Gaussianities of the density distributions have been identified by calculating the third moment of the distributions, i.e. the skewness.

In the more recent literature, an unclear partition is done between the PDF analysis of density and velocity fields. While the global PDFs of star forming regions are generally applied on column density maps \citepads{2010A&A...512A..81F, 2013ApJ...766L..17S}, multiscale PDF analysis in order to detect intermittency is generally reserved to velocity field or centroid velocity increments \citepads{2008A&A...481..367H, 2015MNRAS.446.3777B}. For the turbulent ISM, it is generally assumed that the density PDF follows a lognormal distribution, where the standard deviation  $\sigma$ is related to the Mach number through the relation,

\begin{equation}
\sigma^2 = \ln(1+b^2\mathcal{M}^2),
\label{eq:PDF_Mach}
\end{equation}

\noindent with $b \approx 0.5$ determined by \citetads{1997ApJ...474..730P} with supersonic magnetohydrodynamic (MHD) simulations. On the other hand, as shown by \citetads{1996ApJ...463..623L}, some normalised PDFs of centroid velocity increments have broad to near-exponential wings in PDFs, while global PDFs for the entire velocity field are approximately Gaussian. According to these observations, should the global PDF of turbulent density or velocity fluctuations still be considered as a good measure of the turbulence regime? \citetads{2017A&A...599A..99O} shown on observations, by applying a reconstruction method of the statistical properties of a vector velocity field \citepads{2014MNRAS.442.1451B}, that there can be a strong intra-cloud variability of the compressive and solenoidal fractions. In this case, a more local or scale dependent approach for PDF analyses would more suitable.

In this paper we propose an alternative approach to identify the intermittent behaviour of the ISM, where the non-Gaussianities are identified by calculating the third moment of fluctuation distributions as a function of scales. One powerful aspect of our procedure is the use of complex wavelet transforms, which, contrary to the structure function or Fourier power spectrum analysis, allows us to truly filtered the spatial scales and extract the non-Gaussian component contributing normally to the shallower slope than $\zeta_p = p/3$ for structure functions with $p\geqslant3$. Another advantage of the MnGSeg analysis is that it allows us to calculate a ``non-biased'' power spectrum of a map, i.e. a power spectrum analysis not affected by the non-Gaussianities. For both maps, the column density map and the intensity map at 250 $\mu$m, the Gaussian power spectrum has a power law closer to the Kolmogorov incompressible turbulence of $11/3$. This result could be interpreted, with caution, as a demonstration of the correlation between fluctuations in velocity and density and it shows that the segmentation procedure could extract the incompressible turbulent component. Further analyses applying the MnGSeg technique on centroid velocity maps and numerical simulations, where the intermittent behaviour of both, the density and velocity field, could be compared are needed to confirm this interpretation.

We can also interpret from these results that the $\sim 0.15$ pc `characteristic scale' identified in the ratio between the coherent and the Gaussian power spectra in Fig. \ref{fig:ratio} is associated with the transition of regimes dominated by incompressible turbulence versus compressible modes as well as other physical processes, such as gravity. However, it has been seen in laboratory experiments that incompressible turbulence alone can also produce intermittence. For this reason, we cannot directly attribute the extracted non-Gaussianities to compressive modes only. \citetads{2011A&A...529A...1S} observed many characteristic scales by applying the $\Delta$-variance method over several Galactic clouds. For a $^{13}$CO line-integrated map of Cygnus  X, they revealed two peaks, one at 4 pc and another one at 80 pc. For all the low-mass star formation region they found a double peak structure in their $\Delta$-variance spectrum of their extinstion map. They are localised around 0.4--1.5 pc and 2.9--4.6 pc. In contrat, they found no characteristic scale for Polaris. They stated three main possible causes for these $\Delta$-variance peaks: preferred geometric scales, such as length and width of filaments or clumps, the decaying turbulence scales \citepads{2000A&A...353..339M}, or scales energy injection from external or internal sources.

The origin of the `characteristic scale' identified with the power spectra ratio is thus uncertain. Many mechanisms and other forces in addition to the turbulence are playing a role in the structure formation in the ISM. Among the forces, gravity itself is undoubtedly playing an important role. Coherent structures are also seen in velocity fields. Observations \citepads{2010ApJ...725...17G, 2012A&A...543L...3H, 2017A&A...606A.123H, 2018A&A...610A..77H} and simulations \citepads{2016MNRAS.455.3640S, 2017MNRAS.472..647Z} show that large filamentary structures represent a collection of smaller and coherent subfilaments, sometimes called fibers. According to the simulations, the turbulence being responsible initially for the large-scale density fluctuations, filaments and their substructures formation would then be gravity-driven by accretion. This scenario where dense sub-structures are not anymore linked to the turbulence could be seen to contradict the results of this paper, where the hierarchical nature of the coherent structures seem intimately linked with the energy cascade of the turbulence. This could however suggest that coherent density-driven filaments originate from shocks directly associated to compressive turbulence. Our results are in agreement with the gravoturbulent model of \citetads{1981MNRAS.194..809L}, \citetads{2008ApJ...684..395H} and more recently \citetads{2017ApJ...847..114L} which take into account the scale dependance of the supporting thermal, turbulent, and magnetic energies. The magnetic field is certainly also playing a role in the formation of large filaments, where it has been found oriented perpendicularly to nearby filaments, as Musca and Taurus \citepads{2013A&A...550A..38P, 2016A&A...586A.136P}. In the case of the Polaris flare region, the projected magnetic field was found to be preferentially aligned with the dust filamentary structures \citepads{2015MNRAS.452..715P, 2016MNRAS.462.1517P}. However, even the denser filamentary structure of the Polaris cloud has a lower column density than Musca and Taurus, and by opposition with these two filaments, Polaris flare does not show any sign of star-formation activity. These peculiar properties of Polaris suggest that the cloud is engaged in its initial phases of molecular cloud formation and it could also has an impact on the magnetic field configuration.

\section{Conclusion}\label{sec:conclusion}

The MnGSeg technique is a new powerful analysis method which constitutes a robust alternative to the classical Fourier power spectrum. By coupling the multiscaled PDFs with the power spectrum analysis, this novel technique is sensitive to the progressive contribution of non-Gaussianities towards the small spatial scales. These contributions commonly attributed to turbulence intermittency are usually measured only with higher-order structures function ($p\geqslant4$). The great advantages of the MnGSeg technique, over the Fourier power spectrum or the structure function, are that it can easily expose the relationship between the self-similar Gaussian structures and the progressive contribution of non-Gaussianities as well as it allows the spatial reconstruction of both components.

Using the MnGSeg technique, we prove that the Fourier power spectrum of the Polaris flare is dominated by the power of its denser filamentary structures. The spatial combination of these non-Gaussianities with the fractal scale-free component produces no characteristic scale visible in the Fourier power spectrum. The origin of these non-Gaussianities appearing as dense filamentary structure is likely diverse. Our results are in agreement with a global emerging scenario, also seen through numerical simulations, where turbulence plays a dominant role for the early stages of molecular cloud formation. Then, the interplay of turbulence intermittency, gravity and other mechanisms, such as thermal instability, shocks and the influence of magnetic fields, are quickly responsible for the density enhancement of the medium and ultimately the formation of stellar objects. In this context, the in-depth analysis of this transition between the two regimes, the incompressible random turbulent field, demonstrating a monofractal nature, and the coherent structures, being most likely multifractal and intimately link to the formation of stars, is fundamental for our global understanding of molecular cloud formation and star formation in the ISM. We tentatively measured this transition at a $\sim 0.15$ pc scale, but further tests need to be done to confirm the association between this characteristic scale and the turbulent regime transition.

The MnGSeg technique deserves to be applied also on velocity or centroid velocity maps. However, the direct application on velocity quantities faces mainly three difficulties: the lower spatial resolution of radio line maps, the multiple velocity components along the line of sight, perhaps especially for dense filamentary structures, and the multiple tracers needed to observe the different gas phases and densities.

The future comparison of MnGSeg analyses on multiple column density maps of nearby molecular clouds will allow us to understand the link between the formation of coherent structures in the ISM and the emergence of star formation activity. The application of a fully multifractal analysis of the coherent structures will also help to expose more accurately the plural nature of molecular clouds and to validate if the spatial variations of the turbulence dissipation is ultimately linked to the core mass function of star forming regions.

\section*{Acknowledgements}

The first author wish to thank the precious comments of Isabelle Joncour on this paper. This project has received funding from the European Union’s Horizon 2020 research and innovation programme under the Marie Sk\l odowska-Curie Grant Agreement No. 750920. This research has made use of data from the Herschel Gould Belt survey (HGBS) project (http://gouldbelt-herschel.cea.fr). The HGBS is a Herschel Key Programme jointly carried out by SPIRE Specialist Astronomy Group 3 (SAG 3), scientists of several institutes in the PACS Consortium (CEA Saclay, INAF-IFSI Rome and INAF-Arcetri, KU Leuven, MPIA Heidelberg), and scientists of the Herschel Science Center (HSC). This research also made use of Astropy, a community-developed core Python package for Astronomy \citepads{2013A&A...558A..33A} and its affiliated package APLpy \citepads{2012ascl.soft08017R}.

\bibliographystyle{apj}
\bibliography{biblio}

\end{document}